\pdfoutput=1

\documentclass[12pt,a4paper]{article}
\usepackage{graphicx}  
\usepackage{color}
\usepackage{amsmath} 
\usepackage{amssymb}
\usepackage{amsfonts}
\usepackage{upgreek} 
\usepackage{ifthen} 

\usepackage{ifthen} 
\newboolean{pdflatex}
\setboolean{pdflatex}{true} 
\newboolean{articletitles}
\setboolean{articletitles}{true} 
\newboolean{uprightparticles}
\setboolean{uprightparticles}{true} 
\newboolean{inbibliography}
\setboolean{inbibliography}{false} 

\graphicspath{{./figs/}} 
\usepackage{xspace}
\usepackage{upgreek}

\def\PJ          {\ensuremath{\mathrm{J}}\xspace}                 
\def\PK          {\ensuremath{\mathrm{K}}\xspace}                 
\def\Pp          {\ensuremath{\mathrm{p}}\xspace}                 
\def\Ppsi        {\ensuremath{\uppsi}\xspace}                 
\def\Peta        {\ensuremath{\upeta}\xspace}                 
                 
\def\Ppi         {\ensuremath{\uppi}\xspace}                 
                 
\def\PUpsilon    {\ensuremath{\Upupsilon}\xspace}                 
\def\PW          {\ensuremath{\mathrm{W}}\xspace}                 
\def\PZ          {\ensuremath{\mathrm{Z}}\xspace}                 
\def\PD          {\ensuremath{\mathrm{D}}\xspace}                 
                 
\def\Pb          {\ensuremath{\mathrm{b}}\xspace}                 
\def\Pc          {\ensuremath{\mathrm{c}}\xspace}                 
\def\Ps          {\ensuremath{\mathrm{s}}\xspace}                 
\def\Pg          {\ensuremath{\mathrm{g}}\xspace}                 
\def\PLambda     {\ensuremath{\Uplambda}\xspace}

\newcommand{\tev}       {\ensuremath{\mathrm{\,Te\kern -0.1em V}}\xspace}
\newcommand{\gev}       {\ensuremath{\mathrm{\,Ge\kern -0.1em V}}\xspace}
\newcommand{\mev}       {\ensuremath{\mathrm{\,Me\kern -0.1em V}}\xspace}
\newcommand{\kev}       {\ensuremath{\mathrm{\,ke\kern -0.1em V}}\xspace}
\newcommand{\ev}        {\ensuremath{\mathrm{\,e\kern -0.1em V}}\xspace}
\newcommand{\gevc}      {\ensuremath{{\mathrm{\,Ge\kern -0.1em V\!/}c}}\xspace}
\newcommand{\mevc}      {\ensuremath{{\mathrm{\,Me\kern -0.1em V\!/}c}}\xspace}
\newcommand{\gevcc}     {\ensuremath{{\mathrm{\,Ge\kern -0.1em V\!/}c^2}}\xspace}
\newcommand{\gevgevcccc}{\ensuremath{{\mathrm{\,Ge\kern -0.1em V^2\!/}c^4}}\xspace}
\newcommand{\mevcc}     {\ensuremath{{\mathrm{\,Me\kern -0.1em V\!/}c^2}}\xspace}

\def\mbarn {\ensuremath{\mathrm{ \,mb}}\xspace}

\def\nb    {\ensuremath{\mathrm{ \,nb}}\xspace}

\def\pb    {\ensuremath{\mathrm{ \,pb}}\xspace}
\def\invpb {\ensuremath{\mbox{\,pb}^{-1}}\xspace}
\def\fb    {\ensuremath{\mbox{\,fb}}\xspace}
\def\invfb {\ensuremath{\mbox{\,fb}^{-1}}\xspace}

\def\pion   {{\ensuremath{\Ppi}}\xspace}

\def\pip    {{\ensuremath{\pion^+}}\xspace}

\def\kaon    {{\ensuremath{\PK}}\xspace}
\def\Kp      {{\ensuremath{\kaon^+}}\xspace}
\def\Km      {{\ensuremath{\kaon^-}}\xspace}

\def\sqs                {\ensuremath{\protect\sqrt{s}}\xspace}

\def\proton      {{\ensuremath{\Pp}}\xspace}
\def\antiproton  {{\ensuremath{\bar{\proton}}}\xspace}
\def\jpsi        {{\ensuremath{{\PJ\mskip -3mu/\mskip -2mu\Ppsi\mskip 2mu}}}\xspace}

\def\W           {{\ensuremath{{\PW}}}\xspace}
\def\Z           {{\ensuremath{{\PZ}}}\xspace}
\def\D           {{\ensuremath{{\PD}}}\xspace}

\def\squark      {{\ensuremath{\Ps}}\xspace}
\def\cquark      {{\ensuremath{\Pc}}\xspace}
\def\cquarkbar   {{\ensuremath{\bar\cquark}}\xspace}

\def\bquark      {{\ensuremath{\Pb}}\xspace}

\def\seff        {{\ensuremath{\upsigma_{\mathrm{eff}}}}\xspace}
\def\pt          {{\ensuremath{p_{\mathrm{T}}}}\xspace}
\def\DPS         {{\ensuremath{\mathrm{DPS}}}\xspace}
\def\gluon       {{\ensuremath{\Pg}}\xspace}
\def\Dp          {{\ensuremath{\PD^+}}\xspace}

\def\Dz          {{\ensuremath{\PD^0}}\xspace}
\def\Ds          {{\ensuremath{\PD^+_{\squark}}}\xspace}
\def\Lc          {{\ensuremath{\PLambda^+_{\cquark}}}\xspace}
\def\mumu        {{\ensuremath{\upmu^+\upmu^-}}\xspace}

\def\Charm   {\ensuremath{\mathrm{C}}\xspace}
\def\CC      {\ensuremath{\Charm{}\Charm}\xspace}
\def\psiC    {\ensuremath{\jpsi{}\mathrm{C}}\xspace}

\def\CCbar   {\ensuremath{\mathrm{C}\overline{\mathrm{C}}}\xspace}

\def\DzDz    {\ensuremath{\Dz{}\Dz}\xspace}
\def\DzDp    {\ensuremath{\Dz{}\Dp}\xspace}
\def\DzDs    {\ensuremath{\Dz{}\Ds}\xspace}
\def\DzLc    {\ensuremath{\Dz{}\Lc}\xspace}
\def\DpDp    {\ensuremath{\Dp{}\Dp}\xspace}
\def\DpDs    {\ensuremath{\Dp{}\Ds}\xspace}
\def\DpLc    {\ensuremath{\Dp{}\Lc}\xspace}

\def\psiDz   {\ensuremath{\jpsi{}\PD^0}\xspace}
\def\psiDp   {\ensuremath{\jpsi{}\PD^+}\xspace}
\def\psiDs   {\ensuremath{\jpsi{}\PD^+_{\mathrm{s}}}\xspace}
\def\psiLc   {\ensuremath{\jpsi{}       \PLambda^+_{\mathrm{c}}}\xspace}

\def\deriv   {\ensuremath{\mathrm{d}}}

\def\jpsione {{\ensuremath{{\PJ\mskip -3mu/\mskip -2mu\Ppsi_{1}\mskip 2mu}}}\xspace}
\def\jpsitwo {{\ensuremath{{\PJ\mskip -3mu/\mskip -2mu\Ppsi_{2}\mskip 2mu}}}\xspace}

\def\ptpsi {{\ensuremath{p_{\mathrm{T}}^{\jpsi}}}\xspace}
\def\ypsi  {{\ensuremath{y^{\jpsi}}}\xspace}

\def\seff  {{\ensuremath{\upsigma_{\mathrm{eff}}}}\xspace}

\def\syst  {{\ensuremath{\mathrm{(syst.)}}}}
\def\stat  {{\ensuremath{\mathrm{(stat.)}}}}

\makeindex
\usepackage{multirow}
\usepackage{graphpap}
\usepackage{rotating}
\usepackage{pifont} 
\definecolor{RootOne}  {rgb}{0,0,0}
\definecolor{RootTwo}  {rgb}{1,0,0}
\definecolor{RootThree}{rgb}{0,1,0}
\definecolor{RootFour} {rgb}{0,0,1}
\definecolor{RootFive} {rgb}{1,1,0}
\definecolor{RootSix}  {rgb}{1,0,1}
\definecolor{RootSeven}{rgb}{0,1,1}

\usepackage{cite} 
\usepackage{mciteplus}

\begin{document}

{\normalfont\bfseries\boldmath\huge
  \begin{center}
    Study of double parton 
    scattering processes with heavy quarks\footnote{
      Prepared for: {\it{Multiple Parton Interactions at the~LHC}},
      Eds. P.~Bartalini and J.~R.~Gaunt, World Scientific, 
      Singapore.
    }
  \end{center}
}
\vspace*{0.6cm}
\begin{center}
Ivan Belyaev$^\dag$ and Daria Savrina$^\ddag$\\
Institute for Theoretical and Experimental Physics,\\
B.~Cheremushkinskaya 25, 117218, Moscow, Russia\\  
$^\dag${\tt{Ivan.Belyaev@itep.ru}}, $^\ddag${\tt{Daria.Savrina@cern.ch}}
\end{center}

\begin{abstract}
  Study of double parton scattering processes\,(DPS) involving heavy quarks 
  provides the~most precise probing of factorization hypothesis for DPS for 
  gluon\nobreakdash-mediated processes.
  The~measurements are performed for the~different final states, including 
  open and hidden\nobreakdash-flavour hadrons and for the~different 
  kinematic ranges of incoming gluons.
\end{abstract}



\tableofcontents
\newpage 
\section{Introduction}

The~double parton scattering processes in the~high\nobreakdash-energy 
hadron\nobreakdash-hadron 
interactions attract~\cite{Abramowicz:2013iva,Bansal:2014paa,Astalos:2015ivw,Proceedings:2016tff} 
significant theoretical and experimental interest.
While the~DPS processes first have been observed 
thirty years ago by AFS~collaboration~\cite{Akesson:1986iv}
in the~proton\nobreakdash-proton collisions at relatively low 
energy\linebreak
\mbox{$\sqs=63\gev$}, 
and later studied in  the~$\proton\antiproton$~collisions 
at~\mbox{$\sqs=1.8$} and \mbox{$1.96\tev$} 
by CDF~\mbox{\cite{Abe:1993rv,Abe:1997xk}} and D0~\mbox{\cite{Abazov:2009gc,
  Abazov:2014fha,
  Abazov:2014qba,
  Abazov:2015fbl}} collaborations, 
the~role of DPS~processes  becomes much more
important~\mbox{\cite{Diehl:2011tt,Diehl:2011yj,Szczurek:2015vha}}
at higher energies,
in particular, for 
$\proton\proton$~collisions at~\mbox{$\sqs=7,8$} 
and~\mbox{$13\tev$}~at~Large Hadron Collider\,(LHC).

The cross section of DPS processes~can be expressed as 
\begin{equation}
  \upsigma_{\DPS} = \dfrac{1}{s} \dfrac{\upsigma_{A}\upsigma_{B}}{\seff} \label{eq:pocket}, 
\end{equation}
where $\upsigma_{A,B}$~are the~individual cross sections 
for processes $A$ and $B$,
\mbox{$s=1,2$}~is a~symmetry factor
and $\seff$~is an~effective cross section.
In~the~factorization approach
the~latter is expressed via the~integral 
over the~transverse degrees of freedom of 
the~partons in the~protons and is an~universal
process and scale\nobreakdash-independent constant. 
The~value of \seff is expected to be 
of the~order of the~inelastic cross section, 
but currently it can't be calculated~\cite{Paver:1982yp,
  Calucci:1997ii,
  Calucci:1999yz,
  DelFabbro:2000ds}
from the~basic
principles.
Significant violations~\cite{Korotkikh:2004bz,Gaunt:2010pi}
of the~factorization 
are expected for some kinematic regions and processes, 
{\it{e.g.}}~for the~processes with high\nobreakdash-$x$~partons.

\begin{figure}[t]
  \centering
  \setlength{\unitlength}{1mm}
  \begin{picture}(130,60)
    \put(0,0){
      \includegraphics*[width=65mm,height=60mm,%
      ]{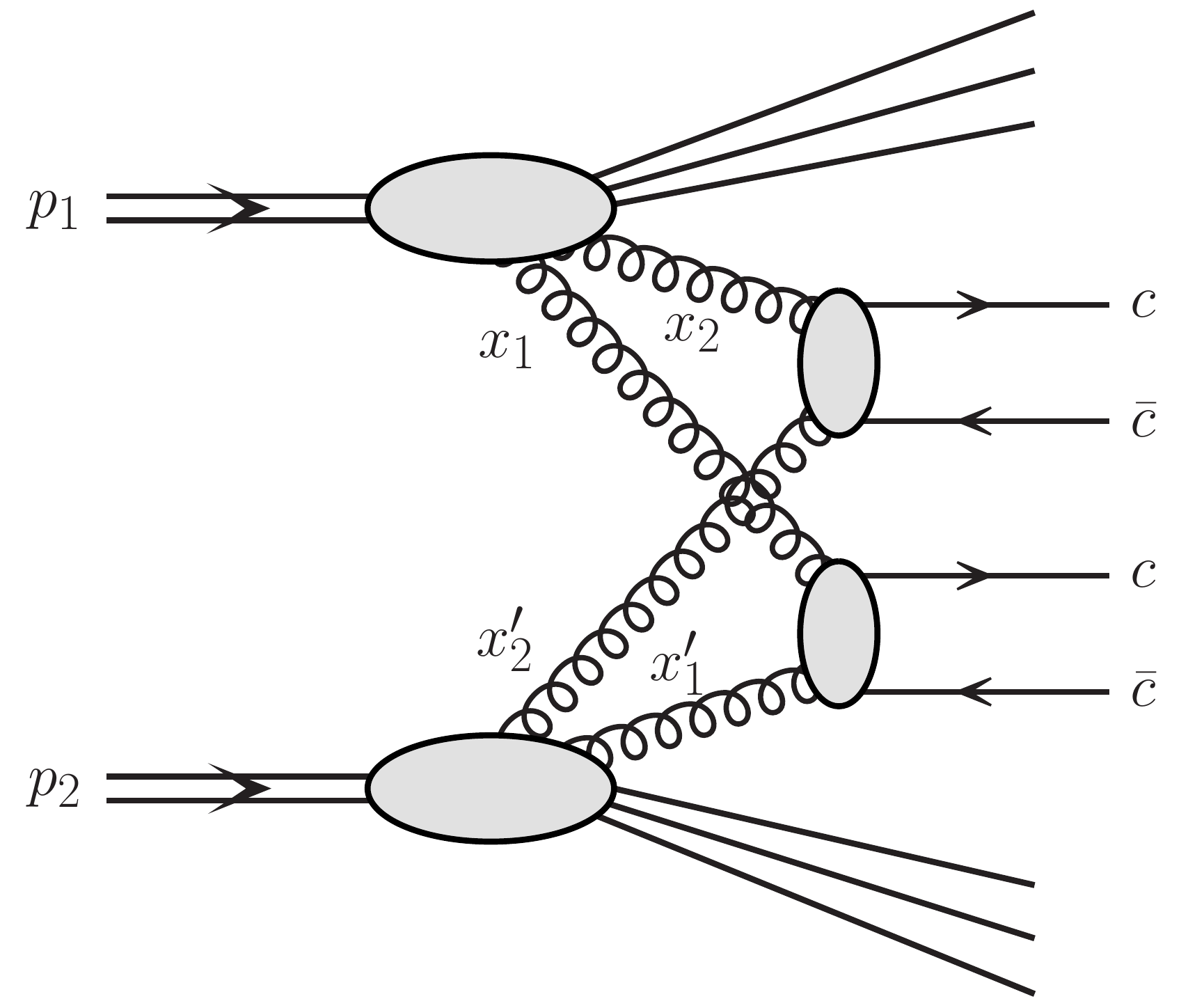}
    }
    \put(65,0){
      \includegraphics*[width=65mm,height=60mm,%
      ]{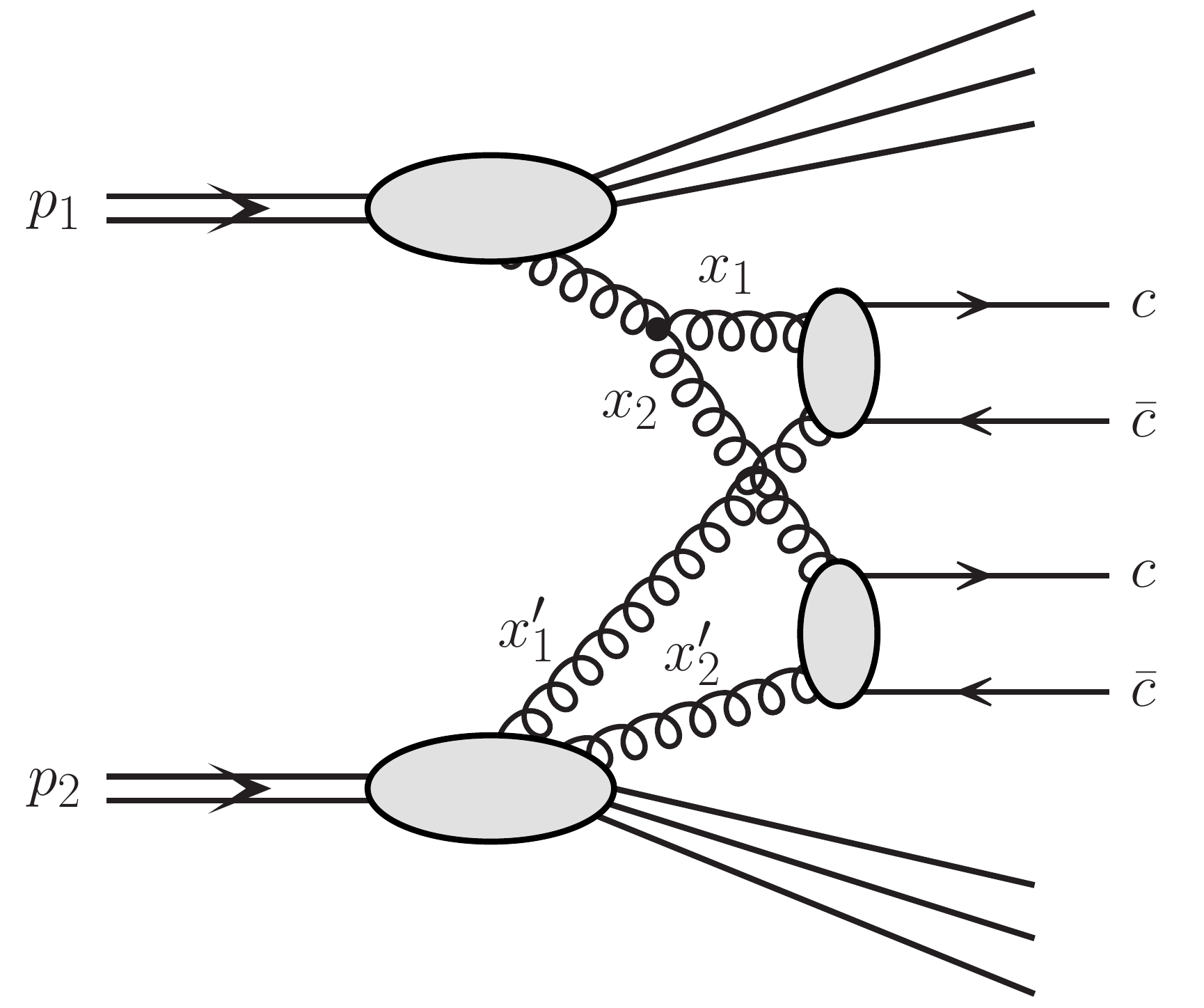}
    }
  \end{picture}
  \caption {\small
    Diagrams for multi\nobreakdash-parton production of 
    $\cquark\cquarkbar\cquark\cquarkbar$:
    $2\otimes2$\,(left) and $1\otimes2$\,(right) processes~\cite{Gaunt:2014rua}.
  }\label{fig:dpshf:fig_one}
\end{figure}

In addition to DPS~$2\otimes2$~process, 
characterized by the~independent scattering 
of two pairs of partons from 
the~incoming protons, Fig.~\ref{fig:dpshf:fig_one}(left), 
one also needs to account for 
the~potentially large additional contribution 
from the~$1\otimes2$~process, 
where a~parton from one incoming proton
splits at some hard scale~\cite{Blok:2010ge,
  Blok:2011bu,
  Blok:2012mw,
  Blok:2013bpa}
and creates two partons 
that further participate in the~two independent 
single\nobreakdash-scattering processes,
see Fig.~\ref{fig:dpshf:fig_one}(right).
This~contribution is calculated~\cite{Blok:2015rka,
  Blok:2015afa,
  Blok:2016lmd}
in pertubative QCD
and provides significant 
dependency of multi\nobreakdash-parton 
interaction cross sections on the~transverse scale.
However for small-$x$ processes with 
relatively small transverse momenta, 
this large dependence is expected 
to stabilize~\cite{Blok:2016ulc}
by accounting of 
soft correlations, turning~\seff to be 
a~constant.

Testing the~universality of \seff, in terms of its 
(in)dependency on the~process, scale and collision energy, 
allows to shed light on the~role of factorization  
and contribution of $1\otimes2$~processes
and provides the~unique opportunity to probe 
the~parton\nobreakdash-parton correlations 
in the~proton. 
Good~understanding of DPS~processes is very important for 
the~search of New Physics\,(NP) effects at~LHC, since for 
certain final states DPS~process mimics~\cite{Krasny:2013aca,Krasny:2015dka,Krasny:2015jka}
the~production 
of heavy exotic particles.
Studies of DPS~processes with heavy\nobreakdash-quarks or quarkonia 
in the~final state are especially important both for QCD tests, 
mentioned above and for the~deep understanding of 
potentially important background source for NP~searches.  

\begin{figure}[t]
  \centering
  \setlength{\unitlength}{1mm}
  \begin{picture}(130,70)
    \put(0,0){
      \includegraphics*[width=65mm,height=70mm,%
      ]{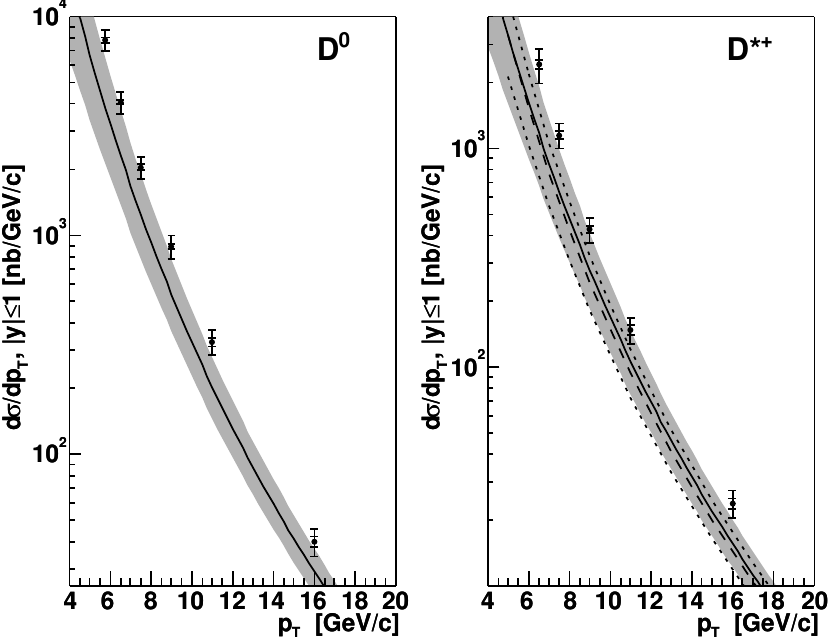}
    }
    \put(65,0){
      \includegraphics*[width=65mm,height=70mm,%
      ]{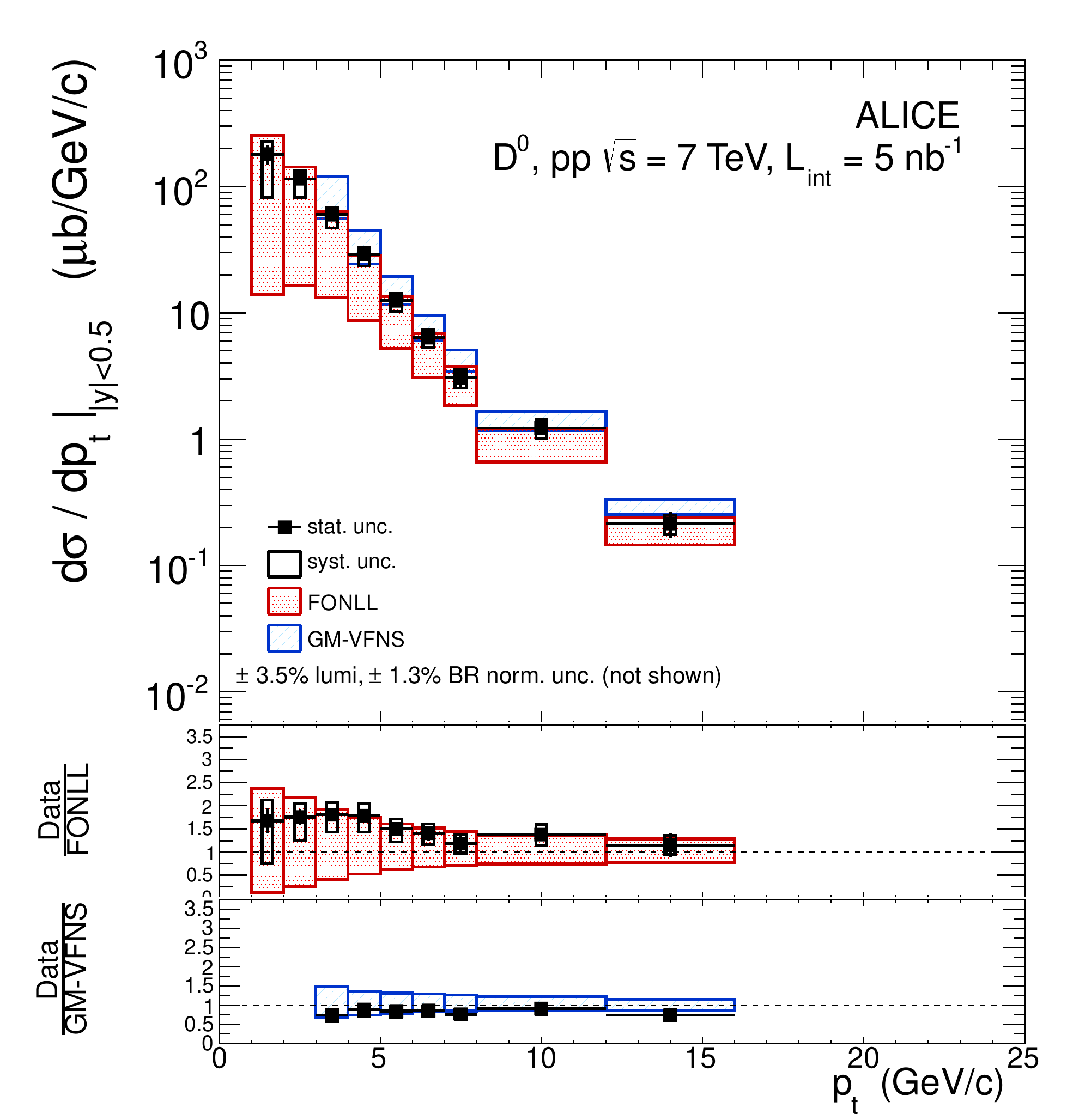}
    }
  \end{picture}
  \caption {\small
    Left:~Inclusive differential production cross sections for 
    prompt \Dz and~$\D^{*+}$~mesons~\cite{Acosta:2003ax}
    in $\proton\antiproton$~collisions at ~\mbox{$\sqrt{s} = 1.96\tev$}.
    The~solid curves are FONLL theoretical predictions 
    with the uncertainties indicated by the shaded bands. The dashed curve
    shown with the~$\D^{*+}$~cross section corresponds to 
    GMVFNS~theoretical prediction; the~dotted lines indicate the~uncertainty. 
    Right:~Inclusive differential production cross section for prompt \Dz~mesons 
    in $\proton\proton$~collisions at $\sqs=7\tev$~\cite{ALICE:2011aa}
    compared with FONLL and GMVFNS theoretical predictions. 
  }\label{fig:dpshf:fig_two}
\end{figure}

For DPS processes producing
four heavy~quarks in the~final state,\linebreak
\mbox{$Q_1\bar{Q}_1Q_2\bar{Q}_2$}, 
the~corresponding elementary subprocesses 
\mbox{$\gluon\gluon\rightarrow Q_1\bar{Q}_1$}
and\linebreak
\mbox{$\gluon\gluon\rightarrow Q_2\bar{Q}_2$} are studied in detail.
The~production of open\nobreakdash-charm hadrons at 
high\nobreakdash-energy hadron collisions 
has been studied in $\proton\proton$~collisions 
by LHCb collaboration~\mbox{\cite{LHCb-PAPER-2016-042,LHCb-PAPER-2012-041,LHCb-PAPER-2015-041}}
 at~\mbox{$\sqs=5$}, 7~and~13\tev,
by~\mbox{ATLAS} collaboration~\cite{Aad:2015zix} 
at~\mbox{$\sqs=7\tev$}
and 
by ALICE~collaboration~\cite{Abelev:2012vra,ALICE:2011aa,Abelev:2012tca}
at~\mbox{$\sqs=2.76\tev$} and \mbox{$7\tev$},
and by  CDF collaboration~\cite{Acosta:2003ax} in 
$\proton\antiproton$~collisions 
at~\mbox{$\sqs=1.96\tev$}.
The~measurements are in reasonable agreement with
calculations at\linebreak
the~next\nobreakdash-to\nobreakdash-leading 
order\,(NLO) using the~generalized mass variable
flavour number scheme~\mbox{\cite{Kniehl:2012ti,Kniehl:2004fy,Kniehl:2005ej,Kneesch:2007ey,Kniehl:2009ar}}\,(GMVFNS),
{\sc{Powheg}}~\cite{Gauld:2015yia}
and fixed order with next\nobreakdash-to\nobreakdash-leading\nobreakdash-log
resummation~\cite{Cacciari:2015fta,Cacciari:1998it,Cacciari:2003zu,Cacciari:2005uk}\,(FONLL),
see Fig.~\ref{fig:dpshf:fig_two}.

Experimentally, the~processes with heavy\nobreakdash-quarks 
often could be studied in details up to very low transverse momenta, 
comparable or even smaller than the~masses of 
the~involved heavy\nobreakdash-quarks,
especially for the~final states with the~dimuon decays of 
the~$\jpsi$ or $\PUpsilon$~mesons. 
It~opens the~unique possibility to explore the~low\nobreakdash-\pt region,  
where DPS~contribution~\cite{Diehl:2011tt,Diehl:2011yj} 
is not suppressed. 
Full reconstruction of the~final state  quarkonia or 
the~open\nobreakdash-flavour heavy hadron allow to minimize 
some large experimental uncertainty, {\it{e.g.}}  related to 
jet reconstuction or jet\nobreakdash-energy calibration.
Currently practically all (experimental) knowledge of 
DPS processes in very interesting and important kinematic region 
\mbox{$x_1,x_2 \gg x_1^{\prime},x_2^{\prime}$} comes from the~measurements 
involving heavy\nobreakdash-quarks~\cite{Aaij:2011yc,Aaij:2012dz,Aaij:2015wpa,Aaij:2016bqq}
in the~forward region, where studies with {\it{e.g.}} light quarks\,(jets) 
and/or the~direct photons are experimentally challenging.

\section{Studies with open-flavour hadrons}\label{sec:dpshf:open_flavour}

\begin{figure}[t]
  \setlength{\unitlength}{1mm}
  \centering
  \begin{picture}(130,70)
    \put(0 ,0){
      \includegraphics*[width=65mm,height=70mm,%
      ]{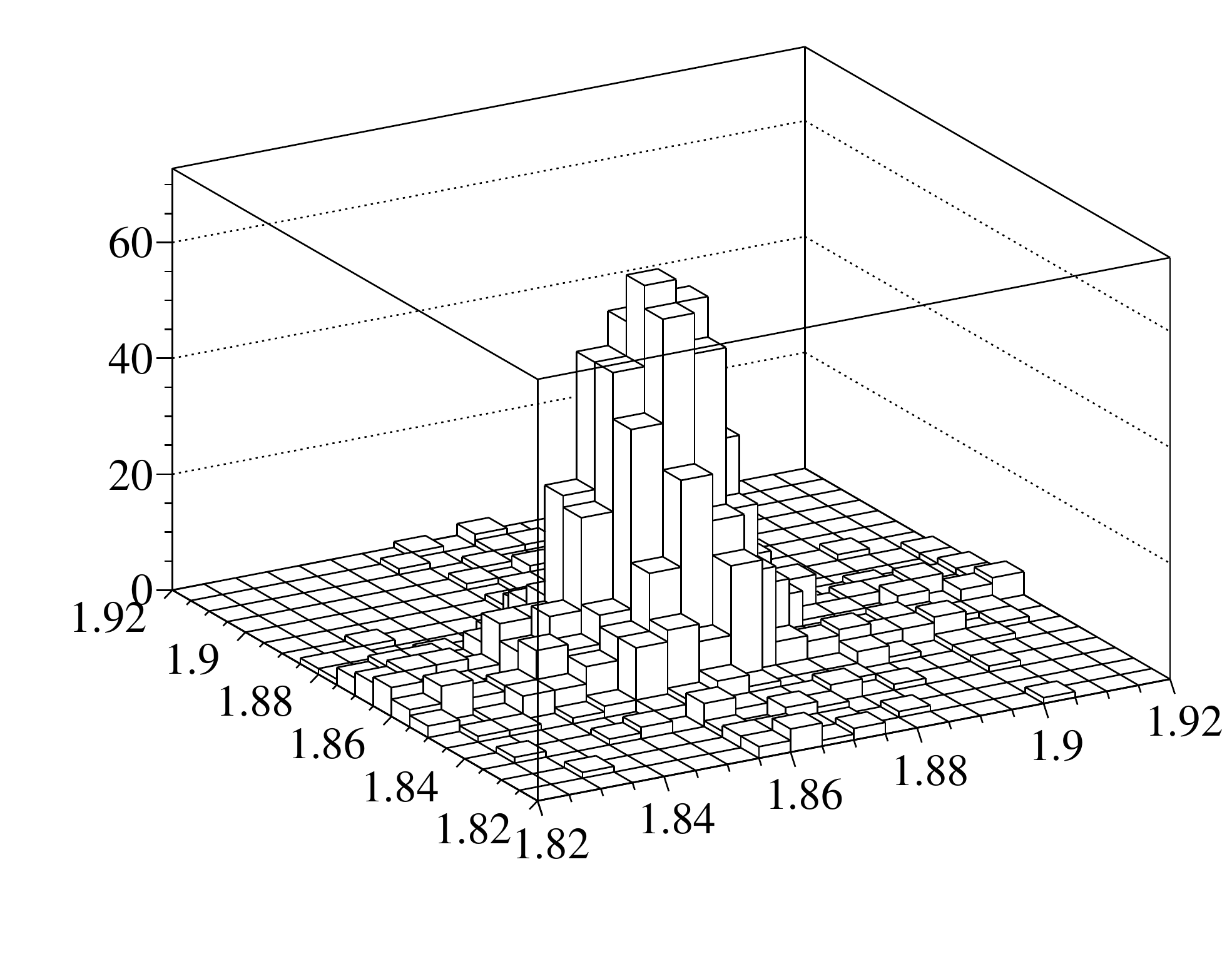}
    }
    \put(65,0){
      \includegraphics*[width=65mm,height=70mm,%
      ]{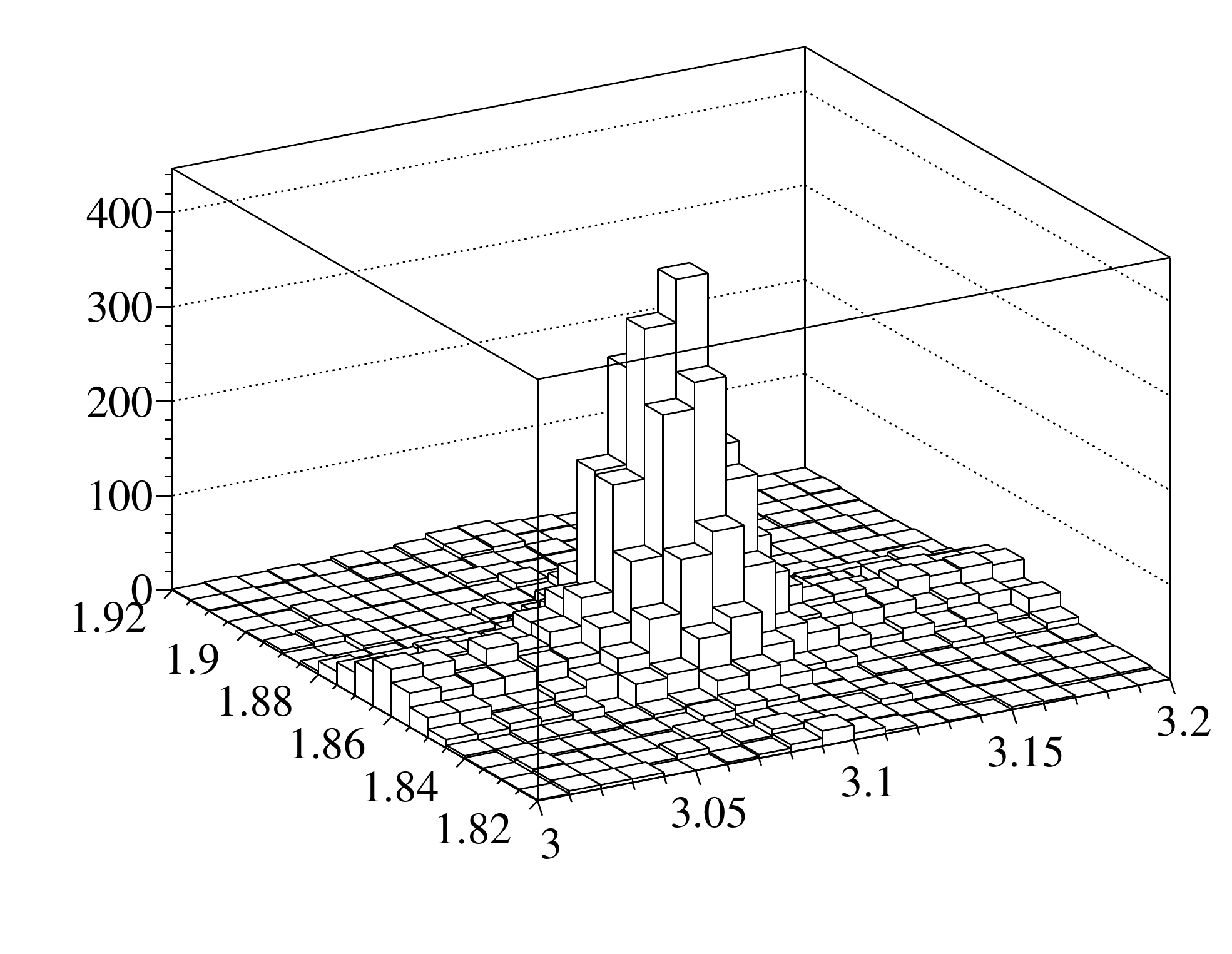}
    }
    \put( 5 , 14)  { 
      \begin{rotate}{-32} \small 
        $m_{\mathrm{K}\Ppi}~\left[\mathrm{GeV}/c^2\right]$
      \end{rotate}
    }
    \put(35,  2)  { 
      \begin{rotate}{10} \small 
        $m_{\mathrm{K}\Ppi}~~~\left[\mathrm{GeV}/c^2\right]$
      \end{rotate}
    }
    \put(100,  4)  { 
      \begin{rotate}{10} \small 
        $m_{\mumu}~~~\left[\mathrm{GeV}/c^2\right]$
      \end{rotate}
    }
    \put(70,  14)  { 
      \begin{rotate}{-32} \small 
        $m_{\mathrm{K}\Ppi}~\left[\mathrm{GeV}/c^2\right]$
      \end{rotate}
    }
    \put(120, 65 ){\small LHCb} 
    \put( 10, 65 ){\small \DzDz} 
    \put( 75, 65 ){\small \psiDz} 
  \end{picture}
  \caption {\small
    Mass distributions for 
    (left)~\DzDz and
    (right)~\psiDz signals.
  }
  \label{fig:dpshf:fig_three}
\end{figure}
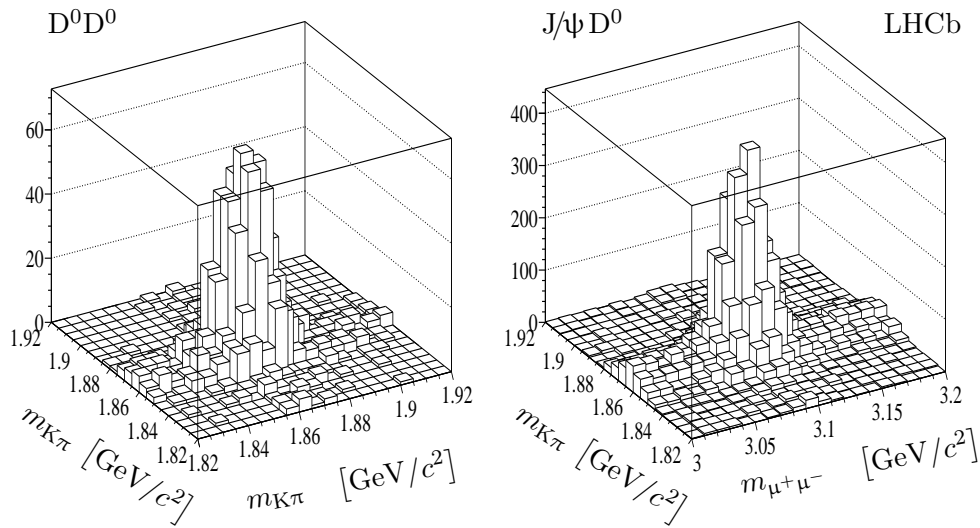

The LHCb experiment studied~\cite{Aaij:2012dz} the~associated production of 
\CC and \psiC~combinations,
where \Charm~stands for \Dz, \Dp, \Ds or \Lc, in the~kinematic region of
\mbox{$2 < \mathrm{y}_{\jpsi}, \mathrm{y_C} < 4.5$},
\mbox{$\mathrm{p_{\jpsi}^T} < 10\gevc$} and
\mbox{$3 < \mathrm{p_C^T} < 12\gevc$}
using \mbox{$355\pm13\invpb$} of data taken in 
$\proton\proton$~collisions at the~centre\nobreakdash-of\nobreakdash-mass 
energy of~\mbox{$\sqs = 7\tev$}.
Open\nobreakdash-charm hadrons are reconstructed via 
\mbox{$\Dz\to\Km\pip$},
\mbox{$\Dp\to\Km\pip\pip$},
\mbox{$\Ds\to\Km\Kp\pip$} and 
\mbox{$\Lc\to\proton\Km\pip$} modes,
while \jpsi~mesons are reconstructed in dimuon final state.
Charge\nobreakdash-conjugated processes are included.
Clear~high\nobreakdash-statistics signals 
with significance in excess of five standard deviations 
have been observed for six \CC~modes,
\DzDz,
\DzDp,
\DzDs,
\DzLc,
\DpDp and 
\DpDs,
and for four \psiC~modes,
\psiDz,
\psiDp,
\psiDs and
\psiLc.
Large~\DzDz and \psiDz~signals are shown in Fig.~\ref{fig:dpshf:fig_three}.
The~possible backgrounds from hadrons produced in two different 
\proton\proton~interactions within  the~same bunch   crossing\,(pile\nobreakdash-up) and contamination 
from \bquark-hadrons decays were reduced to a~negligible level 
by imposing cuts based on the~consistency of the~decay chain.

\begin{figure}[t]
  \setlength{\unitlength}{1mm}
  \centering
  \begin{picture}(130,55)
    \put(  0,0){
      \includegraphics*[height=55mm,width=65mm,%
      ]{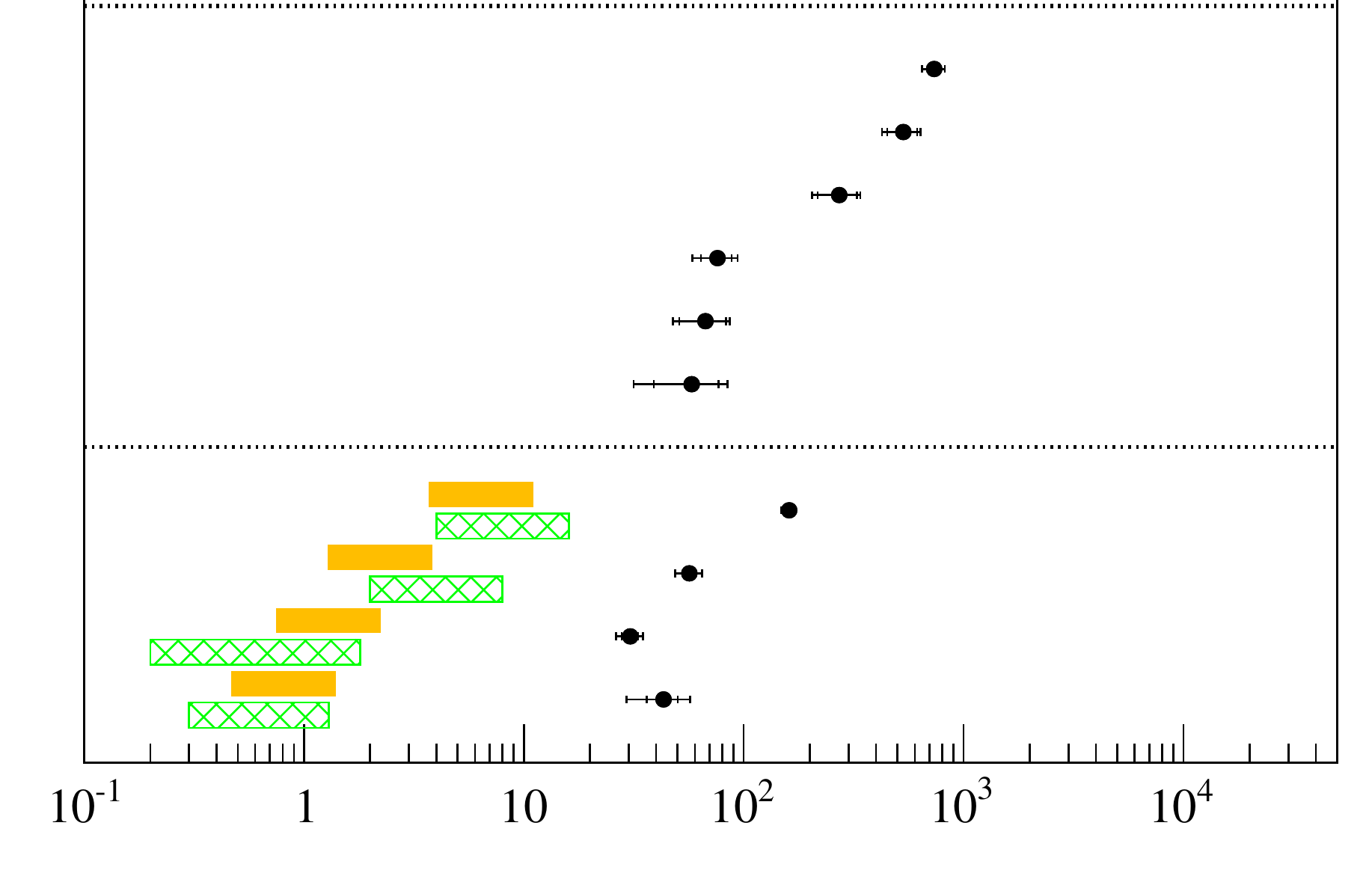}
    }
    \put( 65,0){
      \includegraphics*[height=55mm,width=65mm,%
      ]{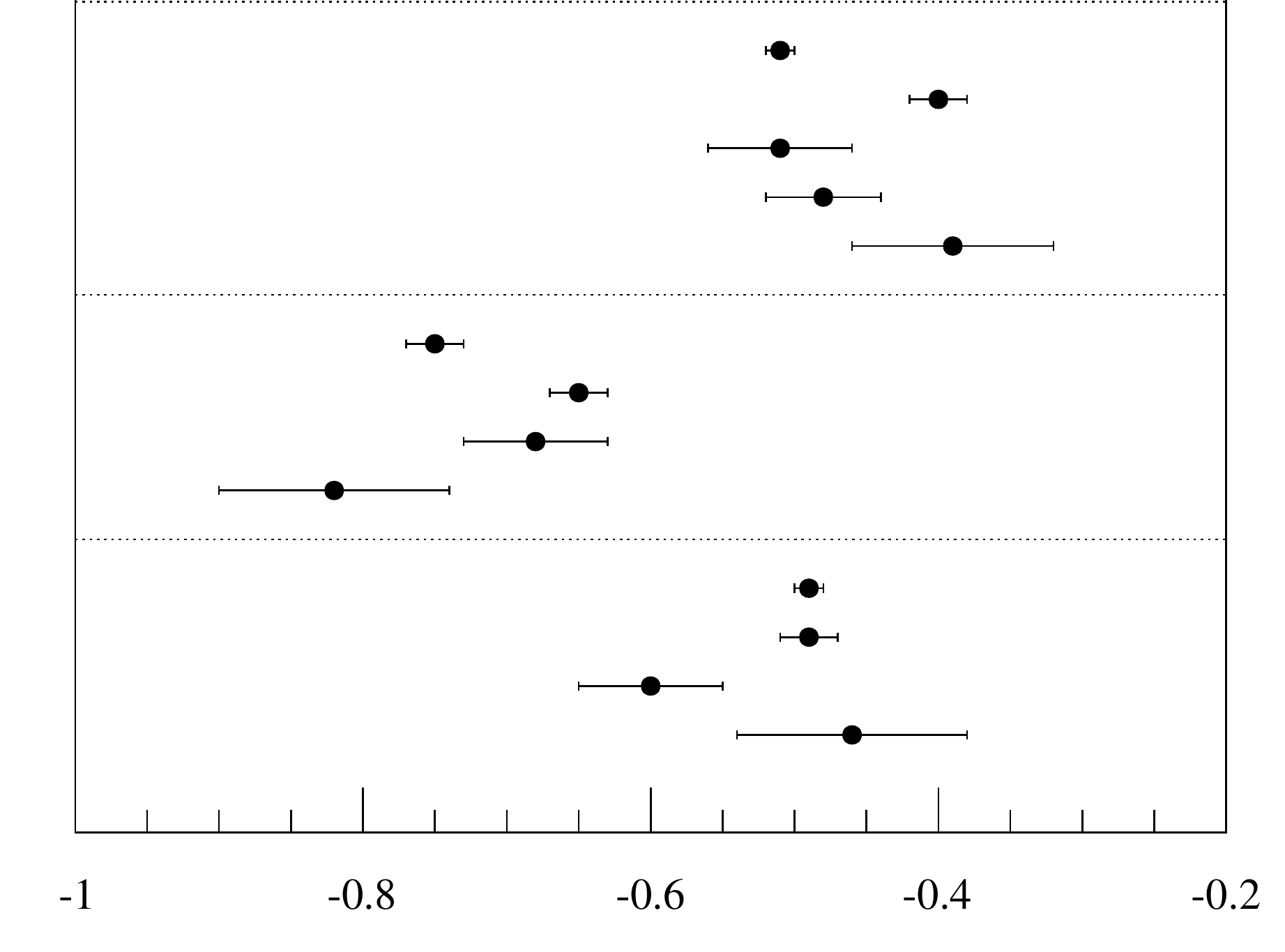}
    }
    \put( 8, 50.0) {\tiny \DzDz  } 
    \put( 8, 46.0) {\tiny \DzDp  } 
    \put( 8, 42.0) {\tiny \DzDs  } 
    \put( 8, 38.0) {\tiny \DpDp  } 
    \put( 8, 34.0) {\tiny \DpDs  } 
    \put( 8, 30.0) {\tiny \DpLc  } 
    \put(55, 22.5) {\tiny \psiDz  } 
    \put(55, 19  ) {\tiny \psiDp  } 
    \put(55, 15.5) {\tiny \psiDs  } 
    \put(55, 12  ) {\tiny \psiLc  }     
    \put(35,  -1 ) {\small  $\upsigma$}
    \put(59,  -1 ) {\small  $\left[\!\nb\right]$}
    \put( 50,50) { \small LHCb}
    \put(115,50) { \small LHCb}
    %
    %
    \put( 80, 51.3) {\tiny \DzDz   } 
    \put( 80, 48.3) {\tiny \DzDp   } 
    \put( 80, 45.3) {\tiny \DzDs   } 
    \put( 80, 42.3) {\tiny \DpDp   } 
    \put( 80, 39.3) {\tiny \DpDs   } 
    \put(110, 34.3 ) {\tiny \psiDz  } 
    \put(110, 31.3 ) {\tiny \psiDp  } 
    \put(110, 28.3 ) {\tiny \psiDs  } 
    \put(110, 25.3 ) {\tiny \psiLc  }     
    \put( 80, 20.0 ) {\tiny \psiDz  } 
    \put( 80, 17.0 ) {\tiny \psiDp  } 
    \put( 80, 14.0 ) {\tiny \psiDs  } 
    \put( 80, 11.0 ) {\tiny \psiLc  }     
    %
    \put(120, 33.0) {\small $p_{\mathrm{T}}^{\Charm}$} 
    \put(120, 19.0) {\small $p_{\mathrm{T}}^{\jpsi}$} 
    \put( 95,-1){\small \pt-slope} 
    \put(116,-1){\small $\left[c/\!\gev\right]$} 
  \end{picture}
  \caption {\small
    (left)~Measured cross sections
    $\sigma_{\CC}$,
    $\sigma_{\CCbar}$~and
    $\sigma_{\psiC}$~(points with error bars).
    For~\psiC~channels, the~SPS theory calculations 
    are shown with hatched~\cite{Berezhnoy:1998aa,Baranov:2006dh}
    and shaded~\cite{Lansberg:2008gk} areas.
    The~inner error bars indicate
    the~statistical uncertainty whilst 
    the~outer error bars indicate
    the~sum of the~statistical and systematic uncertainties in
    quadrature. 
    For~the~\psiC~case the~outermost error bars
    correspond to the~total uncertainties
    including the~uncertainties due to
    the~unknown polarization of the~prompt \jpsi~mesons.
    (right)~The~slope parameters for transverse momentum spectra of charm hadrons 
    for \CC~and \psiC~events.
  }
  \label{fig:dpshf:fig_four}
\end{figure}
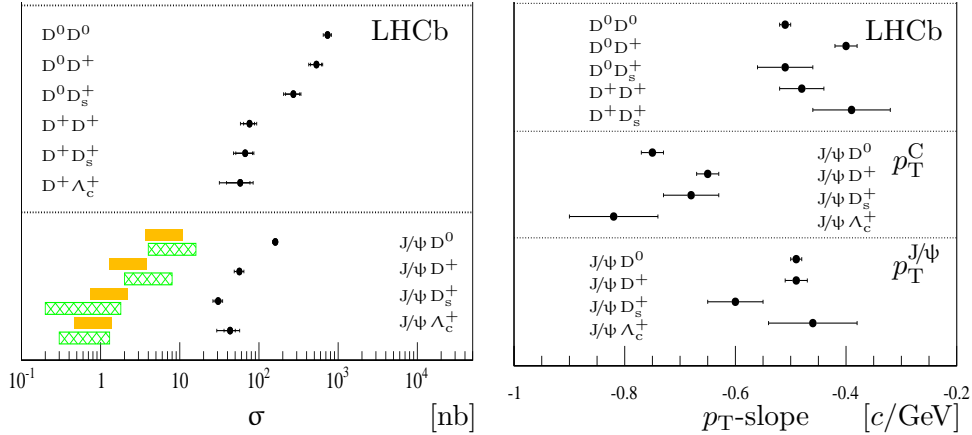

The~model-independent cross sections for double charm production in
the~fiducial range were calculated using the~yields obtained from the
fits of two-dimensional distributions.
The dominating systematic uncertainties are related to 
the~luminosity determination, to the \Charm-hadrons branching 
fractions and to the~differences between
the data and simulated sample used for 
the~reconstruction efficiency determination.
The~resulting cross sections for \CC~and \psiC~events are
shown in the~left part of Fig.~\ref{fig:dpshf:fig_four}
as well as several SPS~theoretical estimations~\cite{Berezhnoy:1998aa,Baranov:2006dh,Lansberg:2008gk}
of \psiC~production.
The~expectations from SPS~process 
are by  a~factor of about~20 below 
the~measured cross sections.
Also~the~SPS predictions 
for \DzDz are also approximately 
30~times smaller~\cite{Luszczak:2011zp,Maciula:2014oya}
than the~measured production rate.
The~observed ratio of cross sections for 
\CC~and corresponding \CCbar~processes, 
proportional to 
$\tfrac{\upsigma(\cquark\cquarkbar\cquark\cquarkbar)}
       {\upsigma(\cquark\cquarkbar)}$, is close to 10\%. 
This value is very large
compared with e.g. those measured 
for $\tfrac{\upsigma_{\jpsi\jpsi}}{\upsigma_{\jpsi}} = (5.1\pm1.0\stat\pm1.1\syst)\times10^{-4}$
in the~same kinematic region~\cite{Aaij:2011yc}.

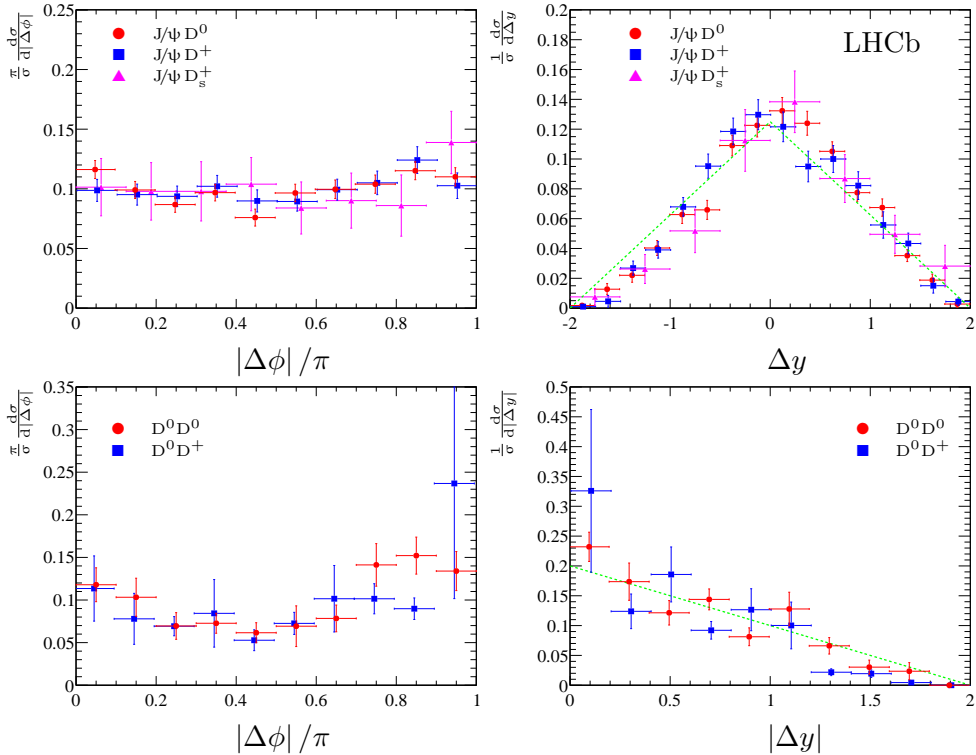
\begin{figure}[t]
  \setlength{\unitlength}{1mm}
  \centering
  \begin{picture}(130,100)
    \put(  0, 50){
      \includegraphics*[width=65mm,height=50mm,%
      ]{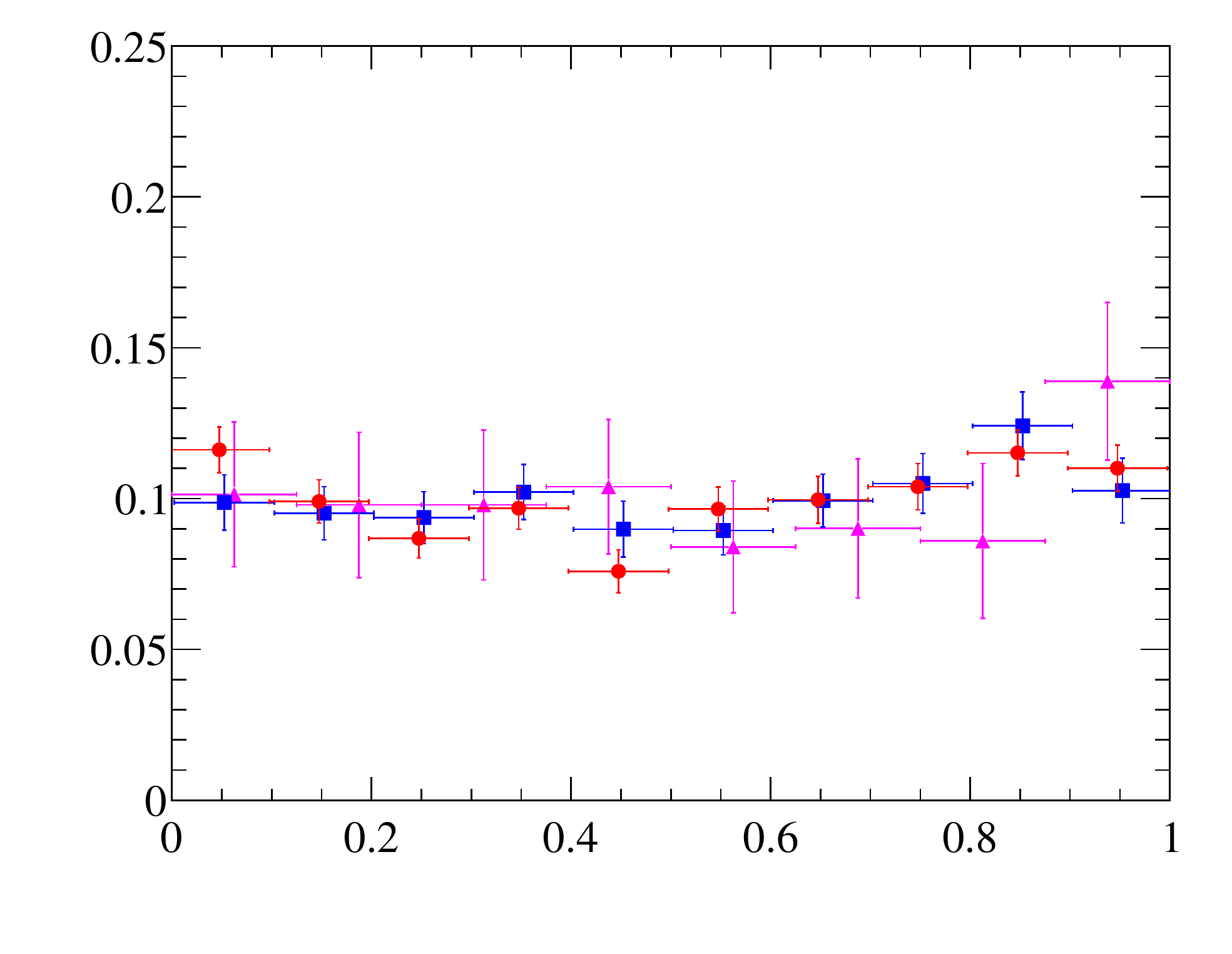}
    }
    \put( 65, 50){
      \includegraphics*[width=65mm,height=50mm,%
      ]{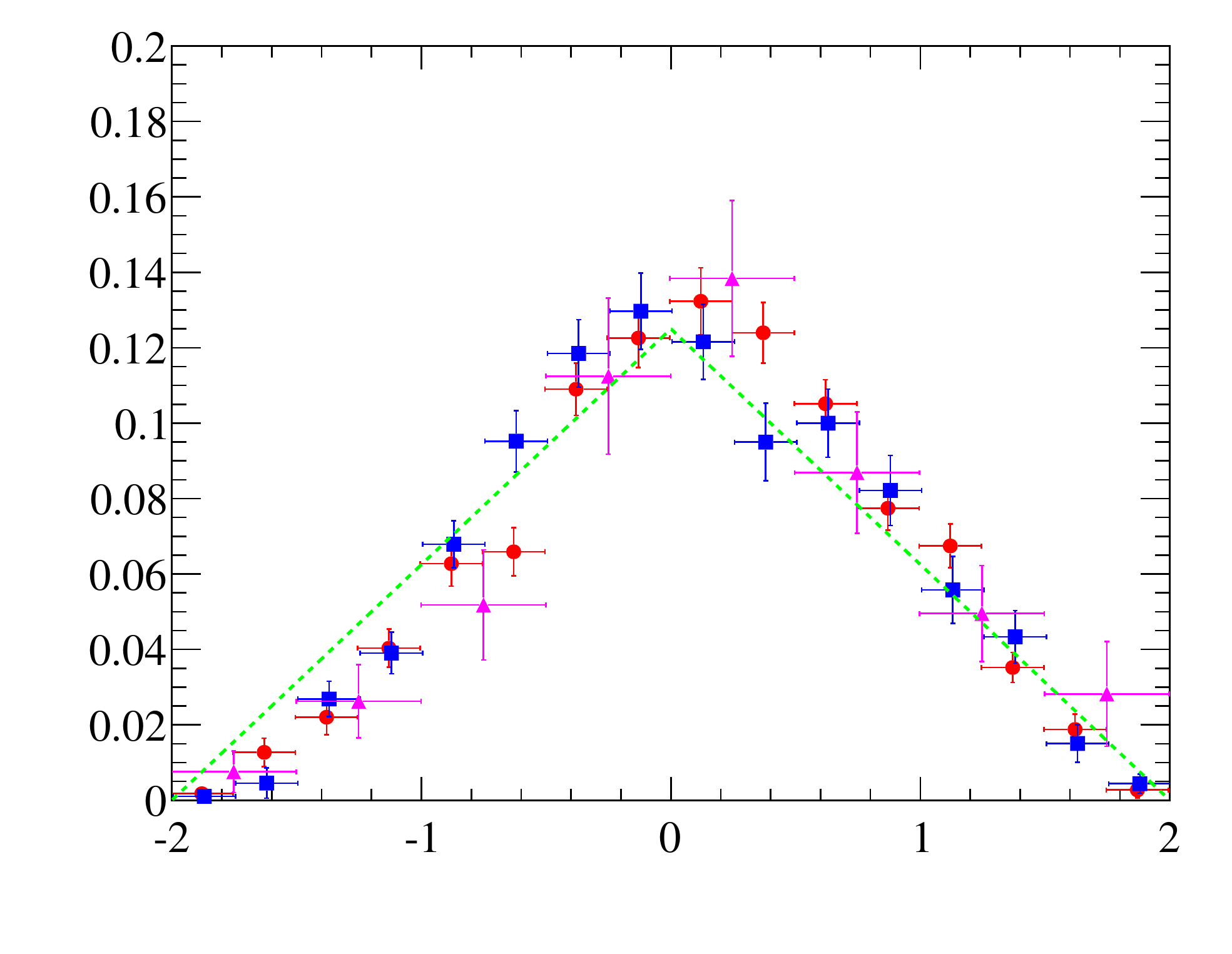}
    }
    \put(  0,  0){
      \includegraphics*[width=65mm,height=50mm,%
      ]{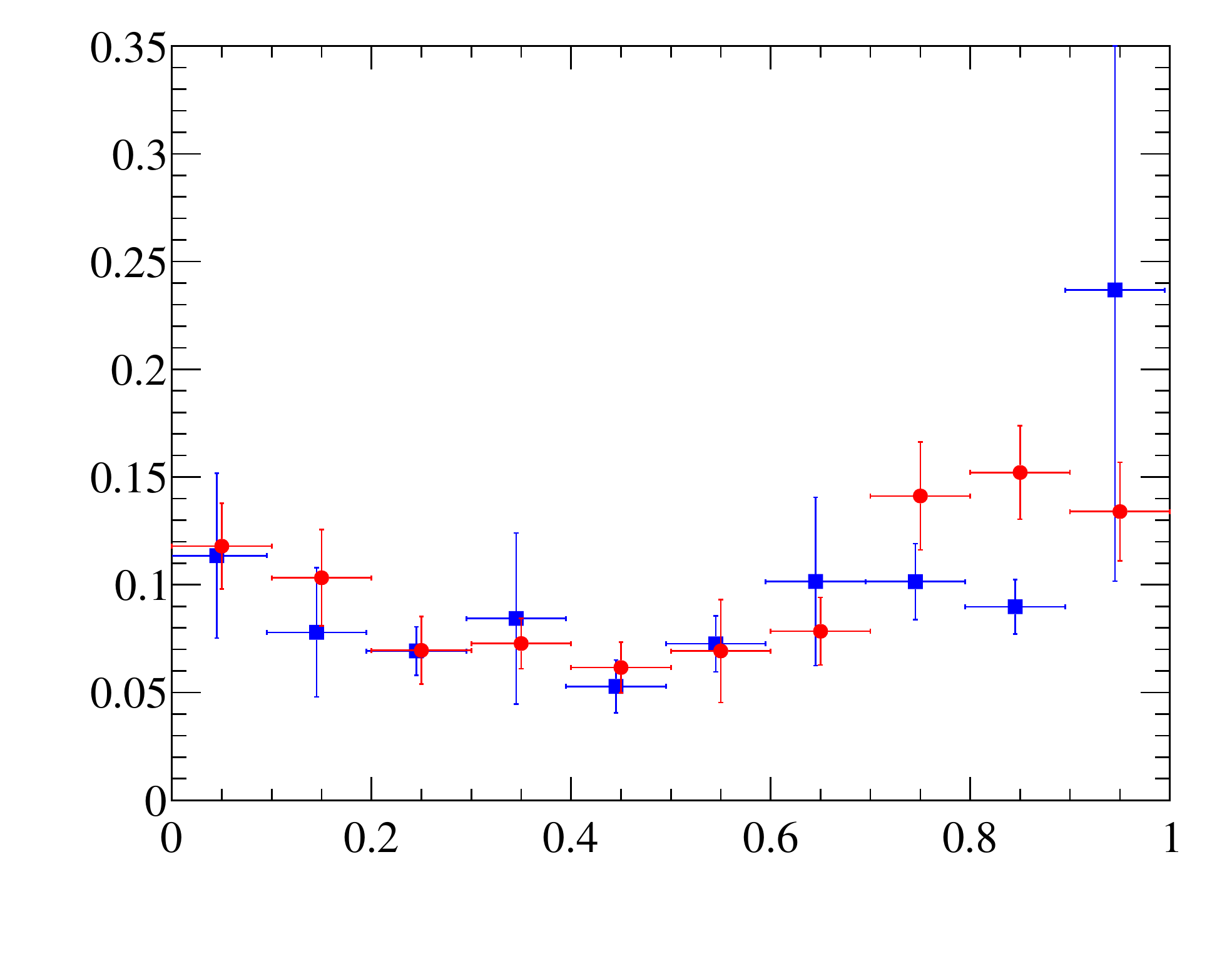}
    }
    \put( 65,  0){
      \includegraphics*[width=65mm,height=50mm,%
      ]{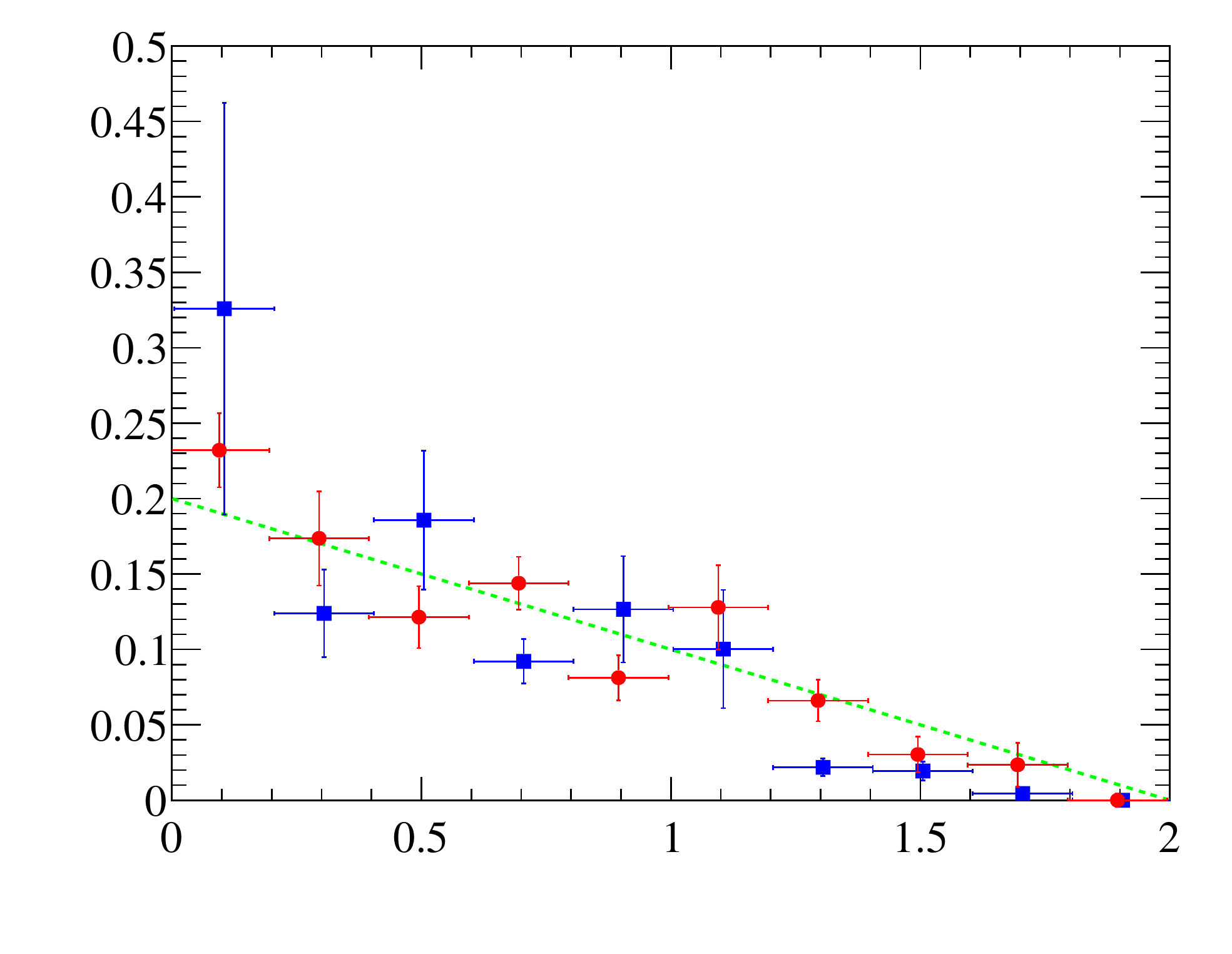}
    }
    \put( 30, 50) { \small $\left|\Delta\phi\right|/\pi$}
    \put(100, 50) { \small $\Delta y$}
    \put(0 , 88){ \tiny  
      \begin{sideways}%
        $\tfrac{\pi}{\upsigma}\tfrac{\deriv\sigma}{\deriv\left| \Delta\phi \right|}$
      \end{sideways}%
    }
    \put(65, 90)  {\tiny  
      \begin{sideways}%
        $\tfrac{1}{\upsigma}\tfrac{\deriv\sigma}{\deriv \Delta y}$
      \end{sideways}%
    }
    \put(12,91){ \tiny
      $\begin{array}{cl}
        {\color{red}     \text{\ding{108}} } & \psiDz  \\
        {\color{blue}    \text{\ding{110}} } & \psiDp  \\
        {\color{RootSix} \text{\ding{115}} } & \psiDs  
      \end{array}$
    } 
    \put(80,91){ \tiny
      $\begin{array}{cl}
        {\color{red}     \text{\ding{108}} } & \psiDz  \\
        {\color{blue}    \text{\ding{110}} } & \psiDp  \\
        {\color{RootSix} \text{\ding{115}} } & \psiDs  
      \end{array}$
    } 
    \put( 30,  0) { \small $\left|\Delta\phi\right|/\pi$}
    \put(100,  0) { \small $\left|\Delta y\right|$}
    \put(0 , 38){ \tiny  
      \begin{sideways}%
        $\tfrac{\pi}{\upsigma}\tfrac{\deriv\sigma}{\deriv\left| \Delta\phi \right|}$
      \end{sideways}%
    }
    \put(65, 38)  {\tiny  
      \begin{sideways}%
        $\tfrac{1}{\upsigma}\tfrac{\deriv\sigma}{\deriv \left|\Delta y\right|}$
      \end{sideways}%
    }
    \put(12,40){ \tiny
      $\begin{array}{cl}
        {\color{red}     \text{\ding{108}} } & \DzDz  \\
        {\color{blue}    \text{\ding{110}} } & \DzDp  
      \end{array}$
    } 
    \put(110,40){ \tiny
      $\begin{array}{cl}
        {\color{red}     \text{\ding{108}} } & \DzDz  \\
        {\color{blue}    \text{\ding{110}} } & \DzDp  
      \end{array}$
    } 
    \put(110,92) { \small LHCb}
  \end{picture}
  \caption {\small
    Distributions of the~difference in azimuthal angle\,(left)
    and rapidity\,(right)
    for \psiC\,(top), \DzDz~and \DzDp~events\,(bottom).
    The~dashed line shows the~expected 
    distribution for uncorrelated events.
  }
  \label{fig:dpshf:fig_five}
\end{figure}

The~kinematic properties of \CC and \psiC~events were studied.
To~compare the~transverse momentum spectra of \jpsi~and open\nobreakdash-charm hadrons, 
the~spectra were fit to an~exponential function.
The~resulting slope parameters are shown in Fig.~\ref{fig:dpshf:fig_four}(right).
The~transverse momentum spectra for \jpsi~mesons 
from \psiC~events are significantly harder than those observed 
in prompt \jpsi~meson production~\cite{Aaij:2011jh}
while the~spectra for open\nobreakdash-charm mesons 
in \psiC~case appear to be very similar to those 
observed~\cite{LHCb-PAPER-2012-041}
for the~prompt charm hadrons.
Similar transverse momenta for \CC~and~\CCbar~events 
are observed. 
However these spectra are much harder than 
those measured~\cite{LHCb-PAPER-2012-041}
for prompt charm events.
It~indicates large correlations between transverse 
momenta of two charm hadrons, possibly 
due to a~large role of 
the~gluon splitting process.
The~rapidity and azimuthal angle distributions, 
presented in Fig.~\ref{fig:dpshf:fig_five},
do not exhibit visible correlations between 
the~two charm hadrons in \CC~and \psiC~events
and are well consistent with uncorrelated production, 
supporting the~dominant role of DPS contribution 
for \CC~and \psiC~production.

Another study exhibiting a~good separation power between 
the~SPS and DPS mechanisms performed by the~LHCb experiment is 
the~study~\cite{Aaij:2015wpa}
of the~associated production of $\PUpsilon$ and
an open\nobreakdash-charm meson.
For~such process
non\nobreakdash-relativistic QCD\,(NR\,QCD)
color\nobreakdash-singlet\,(CS)~\cite{Berezhnoy:2015jga}
and k$_{\mathrm{T}}$\nobreakdash-fac\-to\-ri\-za\-tion~\mbox{\cite{
  Gribov:1984tu,Levin:1990gg,
  Baranov:2002cf,
  Andersson:2002cf,Andersen:2003xj,Andersen:2006pg,
  Baranov:2006dh,Baranov:2006rz,Baranov:2012fb}} 
predict the~ratio 
$\mathcal{R}_{\PUpsilon\cquark\cquarkbar}=\tfrac{\upsigma(\PUpsilon\cquark\cquarkbar)}{\upsigma(\PUpsilon)}$
to be around~\mbox{$(0.1-0.6)\%$}
in LHCb~acceptance, 
while the~DPS predicts this ratio to be of the order of 10\%. 
The measurement was performed with 
the~combined sample 
of data collected
in $\proton\proton$~collisions at 
the~centre\nobreakdash-of\nobreakdash-mass energies
of 7 and 8\tev.
Twelve different
combinations of $\PUpsilon(\mathrm{nS})$\,\mbox{(n=1,2,3)}
and open\nobreakdash-charm hadrons, \Charm, were studied 
in the~fiducial volume of 
\mbox{$2 < y_{\PUpsilon}, y_{\Charm} < 4.5$},
\mbox{$\mathrm{p^{\PUpsilon}_{\mathrm{T}}} < 15~\gevc$} 
and \mbox{$1 < \mathrm{p_{\mathrm{T}}^{\Charm}} < 20~\gevc$}.
In~five of them signal significance exceeds five standard deviations:
$\PUpsilon\mathrm{(1S)}\Dz$,
$\PUpsilon\mathrm{(1S)}\Dp$,
$\PUpsilon\mathrm{(1S)}\Ds$,
$\PUpsilon\mathrm{(2S)}\Dz$ and 
$\PUpsilon\mathrm{(2S)}\Dp$.
Two~of~them with the~highest signal yields,
$\PUpsilon\mathrm{(1S)}\Dz$ and 
$\PUpsilon\mathrm{(1S)}\Dp$~were used for 
the~production
cross section measurements and DPS~studies.
The~ratios $\mathcal{R}_{\PUpsilon\cquark\cquarkbar}$ 
were measured to be $\left(7.7\pm1.0\right)\%$ and 
$\left(8.0\pm0.9\right)\%$ for data sets collected at 
7~and 8\tev, respectively:
these significantly exceed SPS~calculations, 
and agrees with DPS~expectations.

\begin{figure}[t]
  \setlength{\unitlength}{1mm}
  \centering
  \begin{picture}(130,70)
    \put(  0,  0){
      \includegraphics*[width=65mm,height=70mm,%
      ]{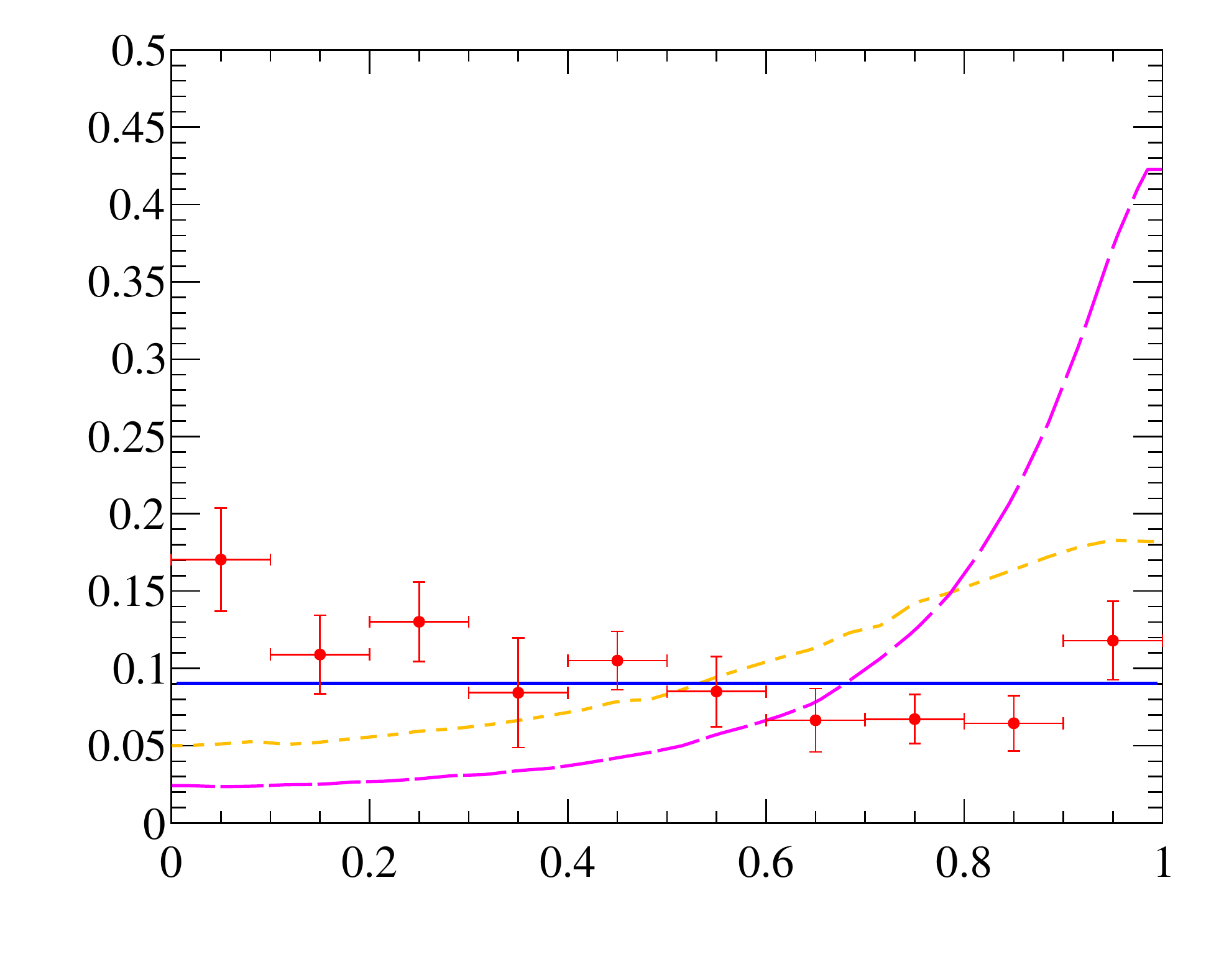}
    }
    \put( 65,  0){
      \includegraphics*[width=65mm,height=70mm,%
      ]{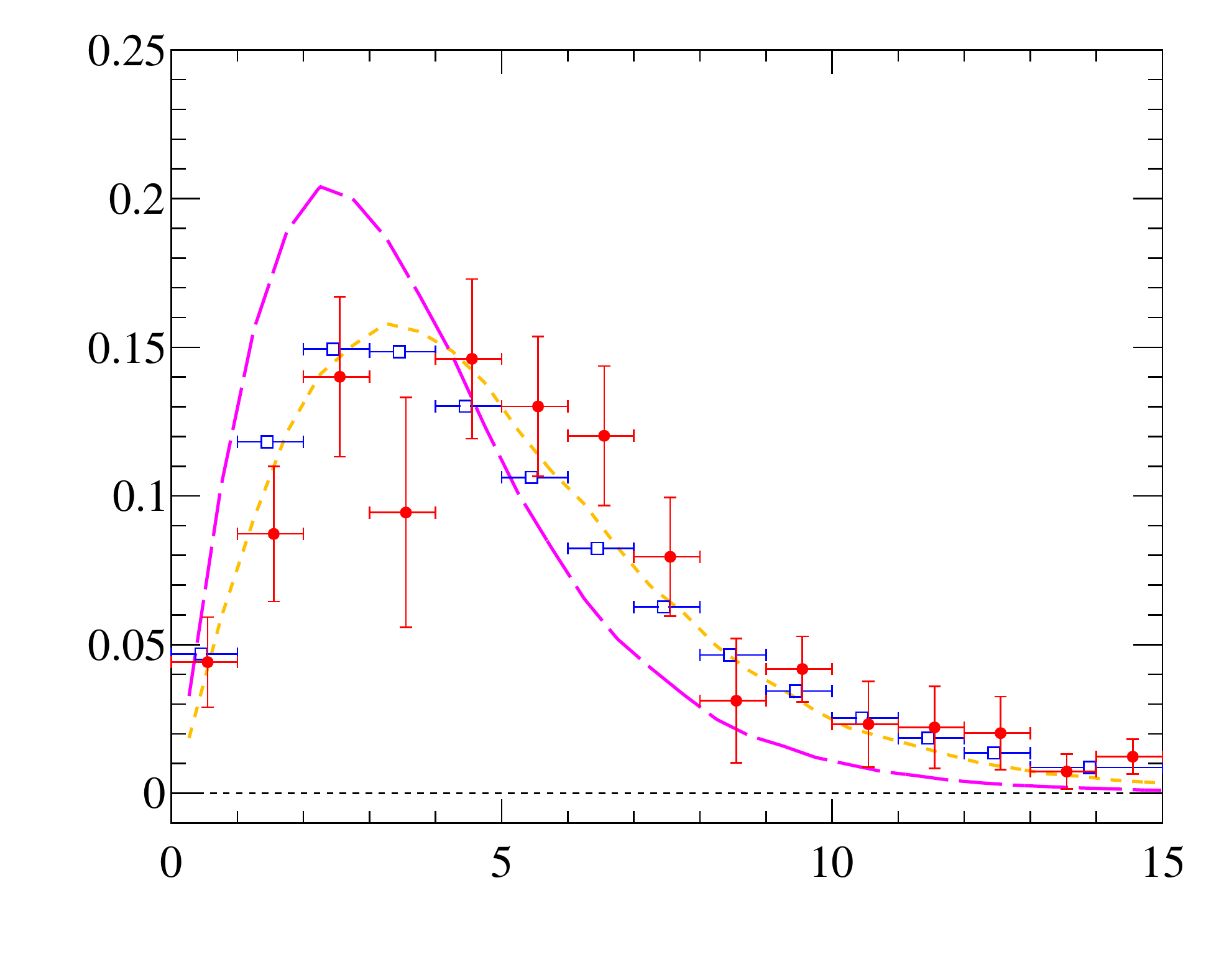}
    }
    \put( 30,  0) { \small $\left|\Delta\phi\right|/\pi$}
    \put(100,  0) { \small $p_{\mathrm{T}}^{\PUpsilon\mathrm{(1S)}}$}
    \put(115,  0) { \small $\left[\!\gevc\right]$}
    \put(0 , 58){ \tiny  
      \begin{sideways}%
        $\tfrac{\pi}{\upsigma}\tfrac{\deriv\sigma}{\deriv\left| \Delta\phi \right|}$
      \end{sideways}%
    }
    \put(65, 58)  {\tiny  
      \begin{sideways}%
        $\tfrac{1}{\upsigma}\tfrac{\deriv\sigma}{\deriv p_{\mathrm{T}}^{\PUpsilon} }$
      \end{sideways}%
    }
    \put(115,60) { \small LHCb}
    \put( 15,60) { \small $\PUpsilon\mathrm{(1S)}\Dz$}
    \put( 80,60) { \small $\PUpsilon\mathrm{(1S)}\Dz$}
  \end{picture}
  \caption {\small
    Distributions for
    $\left|\Delta\phi\right|/\pi$\,(left) and
    $p_{\mathrm{T}}^{\PUpsilon\mathrm{(1S)}}$\,(right) for 
    $\PUpsilon\mathrm{(1S)}\Dz$~events\,(solid red circles).
    Straight blue line in the~$\left|\Delta\phi\right|/\pi$~plots 
    shows the~result of the~fit with a~constant function.
    The~SPS predictions for the~shapes 
    of $\Delta\phi$~distribution 
    are shown 
    with dashed\,(orange) 
    and 
    long\nobreakdash-dashed\,(magenta) 
    curves for calculations based on 
    the~$k_{\mathrm{T}}$\nobreakdash-factorization and 
    the~collinear approximation, respectively.
    The~transverse momentum spectra, 
    derived within the~DPS~mechanism using the~measured
    production cross sections~\cite{LHCb-PAPER-2012-041,Aaij:2015awa} 
    for $\PUpsilon$~and open\nobreakdash-charm mesons,
    are shown with the~open\,(blue) squares.
    All~distributions are normalized to unity.
  }
  \label{fig:dpshf:fig_six}
\end{figure}
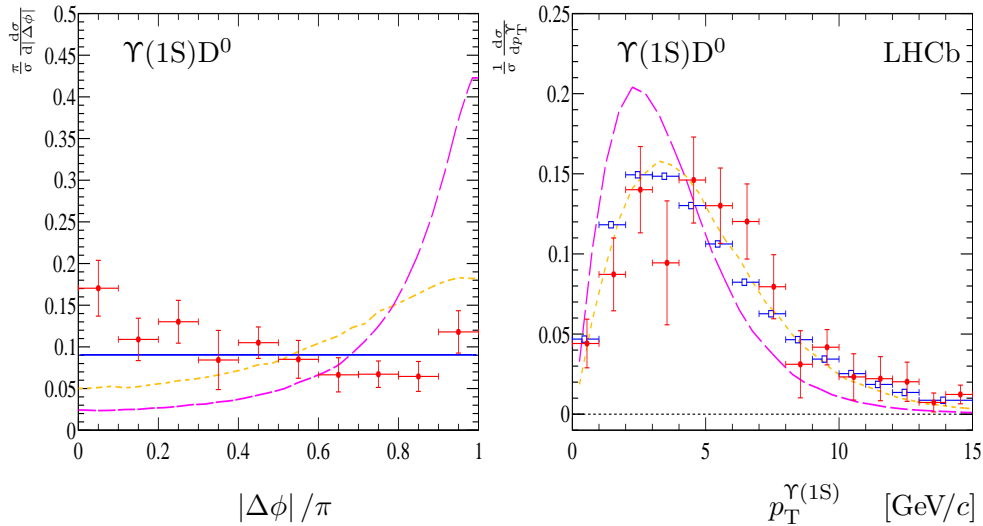

All~differential cross sections for $\PUpsilon\mathrm{(1S)}\D$~events 
nicely agree with expectations from DPS~process.
Figure~\ref{fig:dpshf:fig_six}(left) shows the~distribution 
of the~azimuthal angle difference $\Delta \phi$ for 
$\PUpsilon\mathrm{(1S)}\Dz$~events together 
with both SPS and DPS~expectations. 
The~transverse momentum and rapidity distributions of
$\PUpsilon\mathrm{(1S)}$~mesons also agree well with 
SPS predictions, obtained using 
$k_{\mathrm{T}}$\nobreakdash-factorization approach,
while the~shape of the
transverse momentum spectra of $\PUpsilon\mathrm{(1S)}$~mesons, 
see Fig.~\ref{fig:dpshf:fig_six}(right),
disfavours the~SPS predictions obtained 
using the~collinear approximation.


\section{Double quarkonia studies}\label{sec:dpshf:double_quarkonia}

Unlike the~DPS measurements with open\nobreakdash-flavour hadrons, 
that are unique for the~LHCb experiment, several experiments 
contributed to the~study of the~double quarkonia production. 
The~dimuon decay of quarkonia combines relatively high branching 
fraction and experimentally favourable signature together
with efficient triggering and low background.

First observed by the~NA3 experiment in pion\nobreakdash-nucleon~\cite{Badier:1982ae}
and proton\nobreakdash-nucleon~\cite{Badier:1985ri} interactions,
the~\jpsi~pair production was  later studied both 
in $\proton\antiproton$~collisions at
the~Tevatron by D0~collaboration 
and in $\proton\proton$~collisions 
at~different energies
at the~LHC~\cite{Aaboud:2016fzt,Khachatryan:2014iia,Aaij:2011yc,Aaij:2016bqq}
by ATLAS,
CMS
and LHCb collaborations.

\begin{figure}[t]
 \setlength{\unitlength}{1mm}
 \centering
 \begin{picture}(130,70)
   \put( 0, 0){
     \includegraphics*[width=65mm,height=70mm,%
     ]{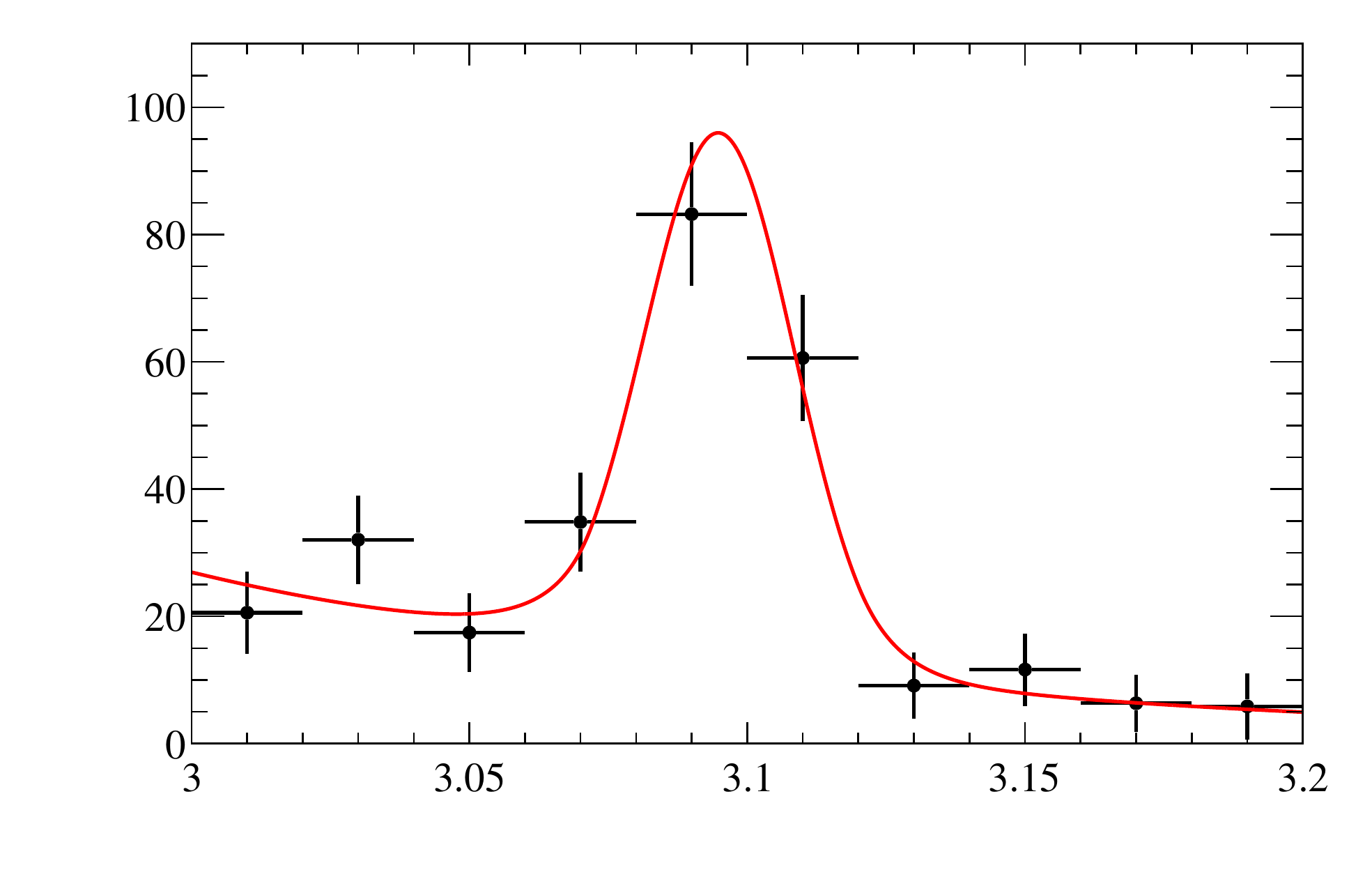}
   }
   \put(65, 0){
     \includegraphics*[width=65mm,height=70mm,%
     ]{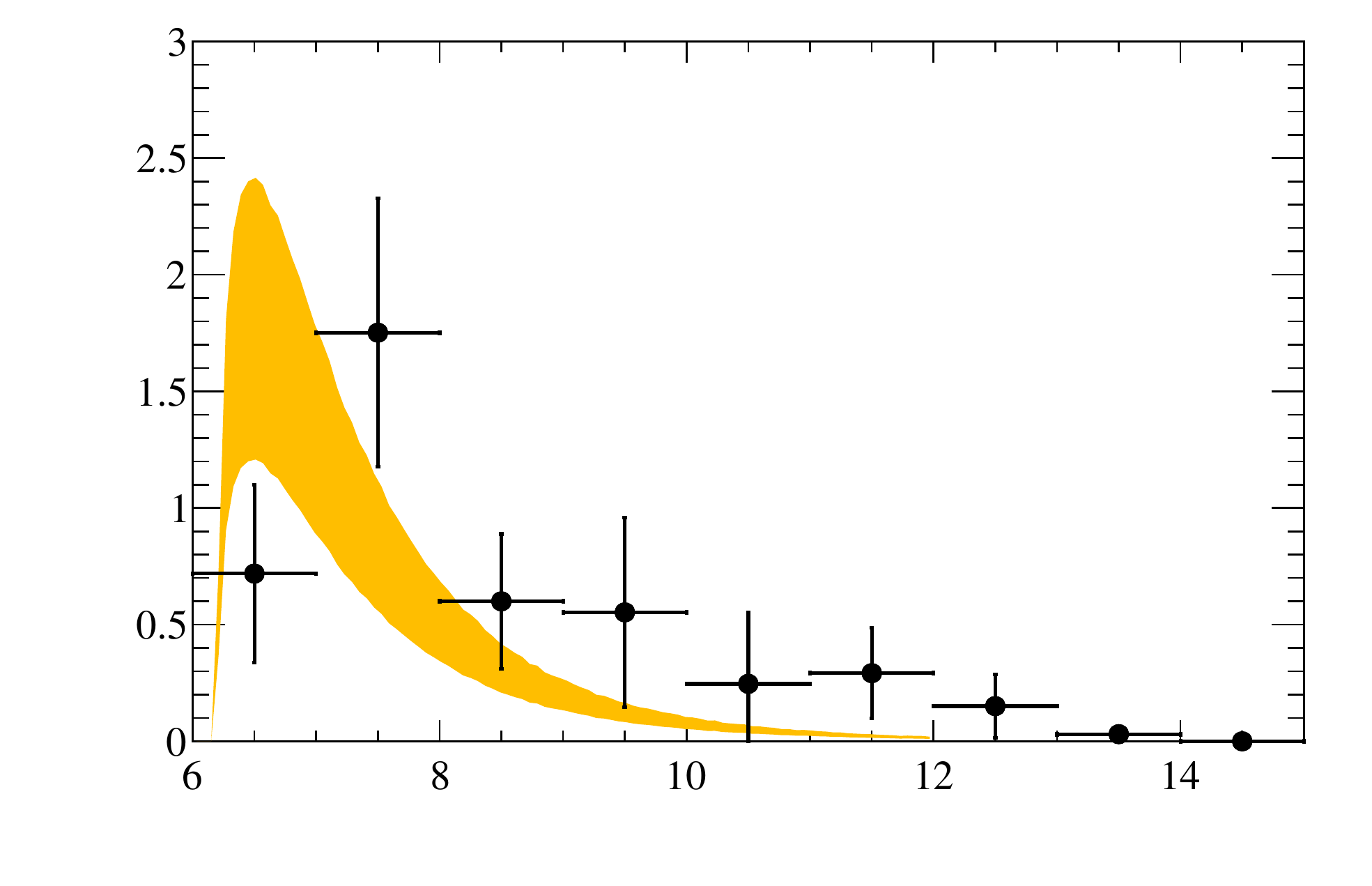}
   }
   \put(0,43) {\tiny
     \begin{sideways}%
       $\frac{ \mathrm{d}N^{\jpsione}}{\mathrm{d}m_{\jpsitwo}} \left[ \frac{1}{20~\mathrm{MeV}/c^2}\right]$
     \end{sideways} 
   }
   \put(65,50) { \tiny
     \begin{sideways}%
       $\frac{\mathrm{d}\sigma}{\mathrm{dm}} \left[ \frac{\mathrm{nb}}{1~\mathrm{GeV}/c^2}\right]$
     \end{sideways} 
   }
   \put( 25, 0) {\small $\mathrm{m}(\mumu)_2$ }
   \put( 50, 0) {\small $\left[\!\gevcc\right]$ }
   \put( 90, 0) {\small $\mathrm{m}^{\jpsi\jpsi}$ }
   \put(115, 0) {\small $\left[\!\gevcc\right]$ }
   \put(115,60){\small LHCb} 
  \end{picture}
 \caption {\small
   (left)\,The fitted yield of $\jpsi \to \left(\mumu\right)_1$
    in bins of $(\mumu)_2$~mass. (right)\,Differential
    production cross section 
    for \jpsi~pairs as a~function of
    mass of the~\jpsi\jpsi~system. 
    The~points correspond to the~data.
    Only~statistical uncertainties
    are included in the~error bars. 
    The~shaded area corresponds
    to LO\,CS predictions~\cite{Berezhnoy:2011xy}.
 }
 \label{fig:dpshf:fig_seven}
\end{figure}
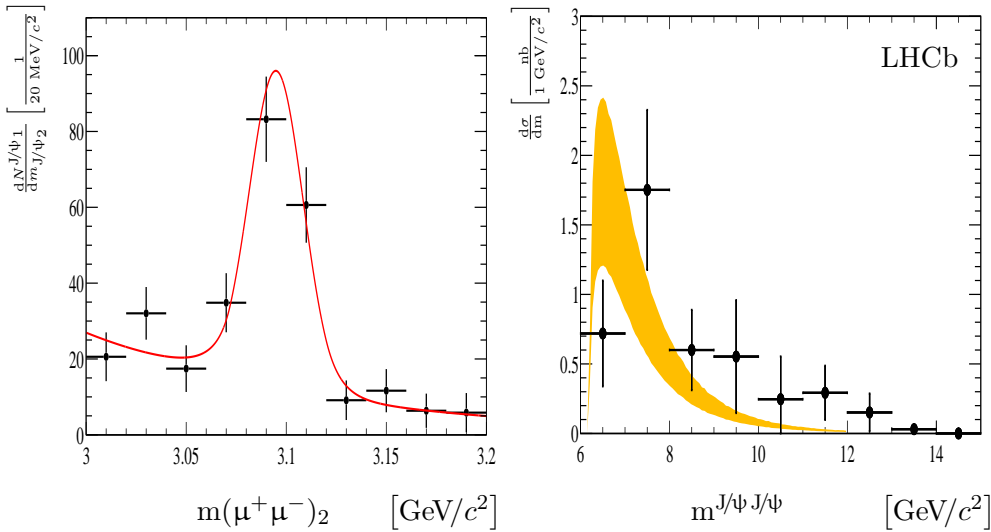

The~first measurement in the~$\proton\proton$~collisions~\cite{Aaij:2011yc}
was made by the~LHCb experiment.
The~oppositely charged tracks identified as muons were combined 
to obtain pairs of \jpsi~candidates\,($\jpsione$ and $\jpsitwo$).
To~determine the~signal event 
yield the~invariant mass of the~second muon pair 
was plotted in bins of the~first pair invariant mass, 
see Fig.~\ref{fig:dpshf:fig_seven}(left). 
Fit~to this distribution was 
performed with a~function including a~signal component for 
\jpsi and a~component for the~combinatorial background.
Other~sources of background were found to be negligible.
In~total 
$141\pm19$ \jpsi pairs were found with the signal significance 
exceeding 6~standard deviations.

The~total cross section in 
the~region~\mbox{$2<\ypsi<4.5$},
\mbox{$\ptpsi<10\gevc$}
was found to be
\begin{equation*}
\upsigma^{\jpsi\jpsi}=5.1\pm1.0\stat\pm1.1\syst\,\mathrm{nb}.
\end{equation*}
This~value is not precise enough to
distinguish~\cite{Baranov:2011ch,Baranov:2012re}
between the~SPS and DPS contributions.
The~SPS contributions are calculated 
to be 
$4.0\pm1.2\nb$ and $4.6\pm1.1\nb$
in the~leading\nobreakdash-order~\cite{Berezhnoy:2011xy,Qiao:2009kg,Sun:2014gca} 
NR\,QCD\,CS approach,
and $5.4^{+2.7}_{-1.1}\nb$
using complete~\cite{Sun:2014gca}
NLO\,CS~approach.
The~DPS contribution is expected 
to be $3.8\pm1.3\nb$.
The~differential production cross section as 
a~function of 
the~mass of
the~$\jpsi\jpsi$~system together with 
expectations from leading\nobreakdash-order\,(LO)\,CS~calculations
is presented in~Fig.~\ref{fig:dpshf:fig_seven}(right).
The~statistics didn't allow to study 
the~kinematic properties of these events and make 
firm conclusions.

\begin{figure}[t]
  \setlength{\unitlength}{1mm}
  \centering
  \begin{picture}(130,70)
    \put( 0, 0){
      \includegraphics*[width=65mm,height=70mm,%
      ]{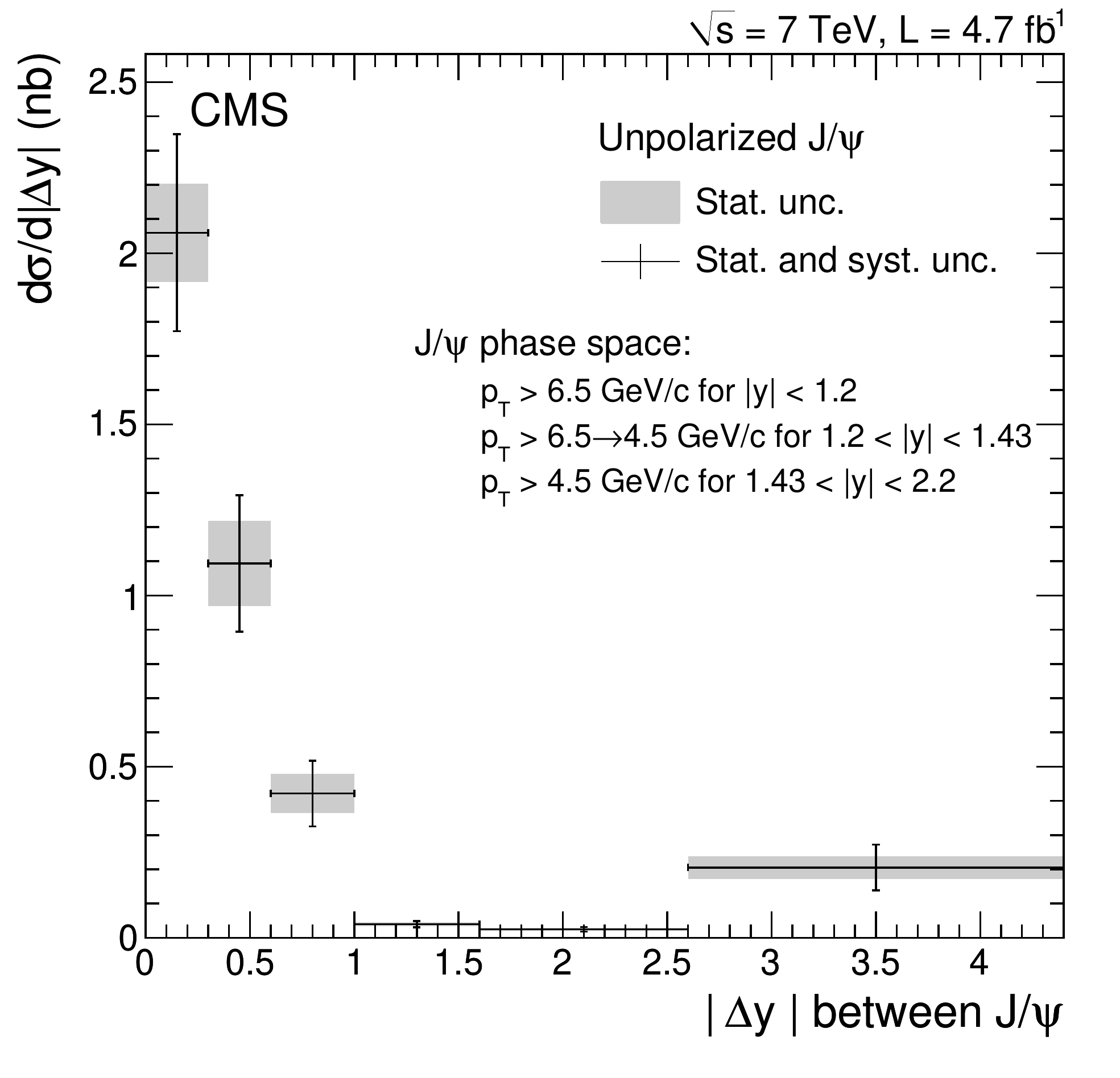}
    }
    \put(65, 0){
      \includegraphics*[width=65mm,height=70mm,%
      ]{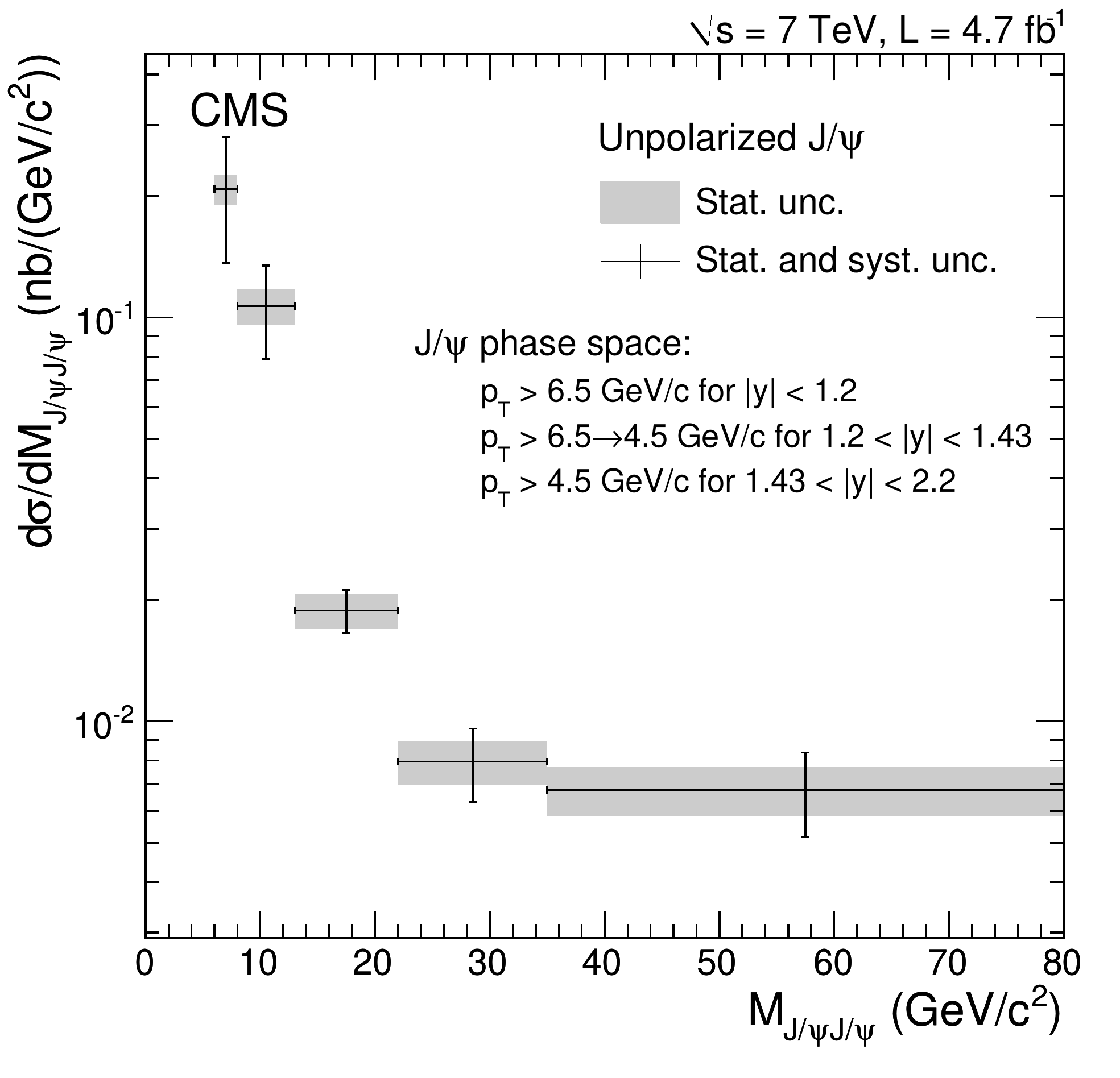}
    }
 \end{picture}
  \caption {\small
    Differential cross section for prompt \jpsi~pair production
   as a~function of 
   the~absolute rapidity difference between \jpsi~mesons\,(left)
   and the~mass of \jpsi\jpsi~system\,(right)
   The shaded regions represent the statistical uncertainties only, 
   and the~error bars
   represent the~statistical and systematic uncertainties 
   added in quadrature.
 }
 \label{fig:dpshf:fig_eight}
\end{figure}

Another measurement at~\mbox{$\sqs=7\tev$} was 
performed~\cite{Khachatryan:2014iia}
by the~CMS collaboration,
which included the~kinematic regions 
where the~color\nobreakdash-octet\,(CO) models could play
a~more significant role in the~double heavy quarkonium production.
After~excluding possible sources of background,
they observed \mbox{$446\pm23$}~\jpsi~pairs produced 
promptly in the~same \proton\proton~collision.
The~cross section in the~fiducial region was measured to be
\begin{equation*}
  \upsigma^{\jpsi\jpsi}=1.49\pm0.07\stat\pm0.13\syst\nb.
\end{equation*} 
The~differential cross sections as 
a~function of the~rapidity difference
between the~two \jpsi~mesons, 
$\left|\Delta y\right|$,
and $M_{\jpsi\jpsi}$, the~mass of 
\jpsi\jpsi~system, 
 are shown in Fig.~\ref{fig:dpshf:fig_eight}.
These~distributions are very sensitive to 
the~DPS effects~\mbox{\cite{Baranov:2012re,Baranov:2011ch,Kom:2011bd,Lansberg:2014swa}}
and non\nobreakdash-vanishing cross sections 
for large $\left| \Delta y \right|$ and $M_{\jpsi\jpsi}$~bins could be 
considered as a~sign of DPS mechanism.

These~distributions have been analyzed~\cite{Lansberg:2014swa}  
against incomplete NLO$^{\ast}$\,CS predictions. 
From~this analysis Lansberg\&Shao~\cite{Lansberg:2014swa}  
concluded the~importance of $\upalpha_{\mathrm{s}}^5$ contributions 
at medium and large transverse momenta
and  dominance of DPS mechanism at large 
$\left| \Delta y \right|$ and large  $M_{\jpsi\jpsi}$.
No~significant CO~contribution was found. 
The~extracted value of 
\seff~parameter has been found~\cite{Lansberg:2014swa} 
to be $\seff=11.0\pm 2.9\mbarn$.

\begin{figure}[t]
  \setlength{\unitlength}{1mm}
  \centering
  \begin{picture}(130,70)
    \put( 0, 0){
      \includegraphics*[width=65mm,height=70mm,%
      ]{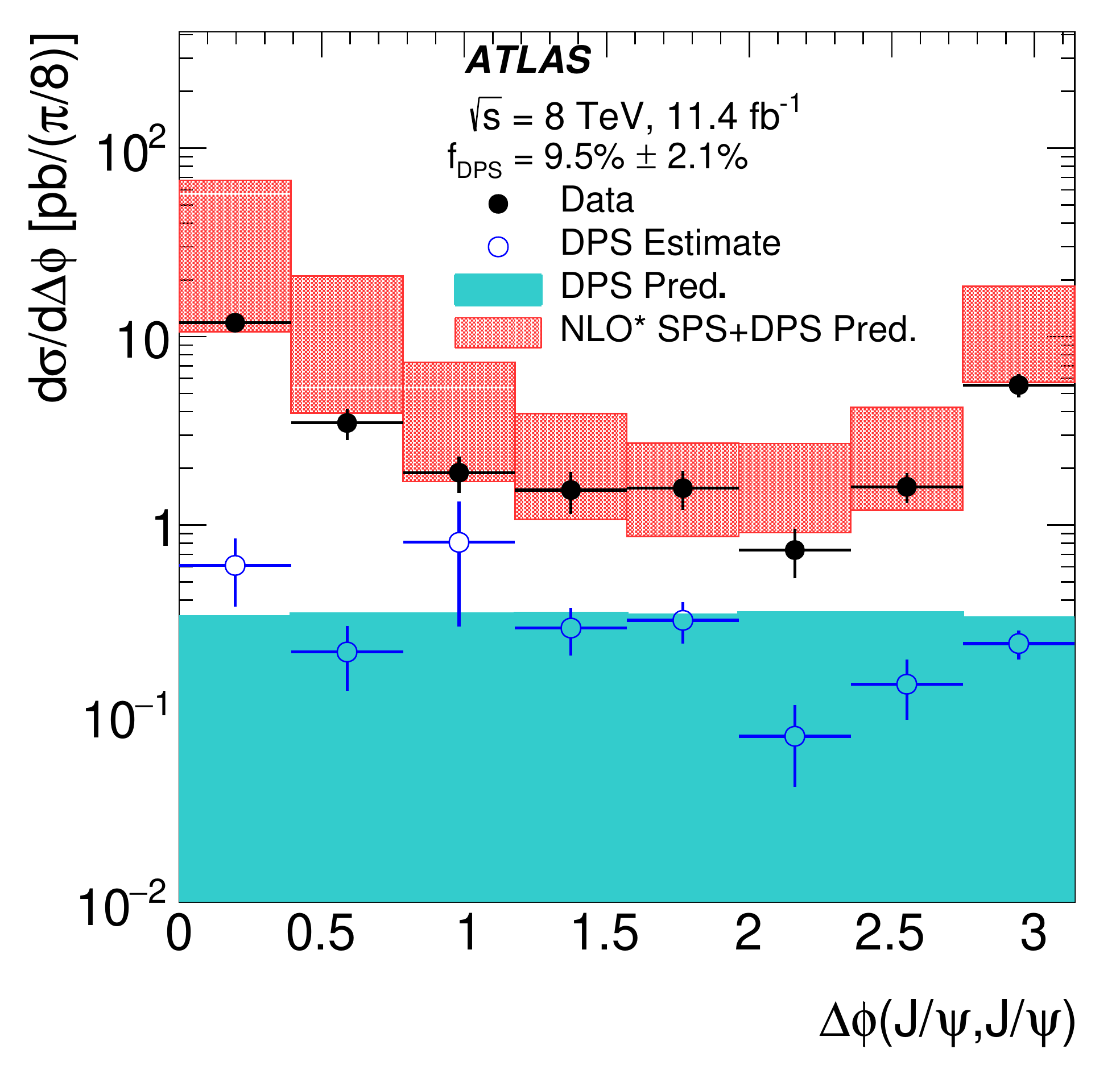}
    }
    \put(65, 0){
      \includegraphics*[width=65mm,height=70mm,%
      ]{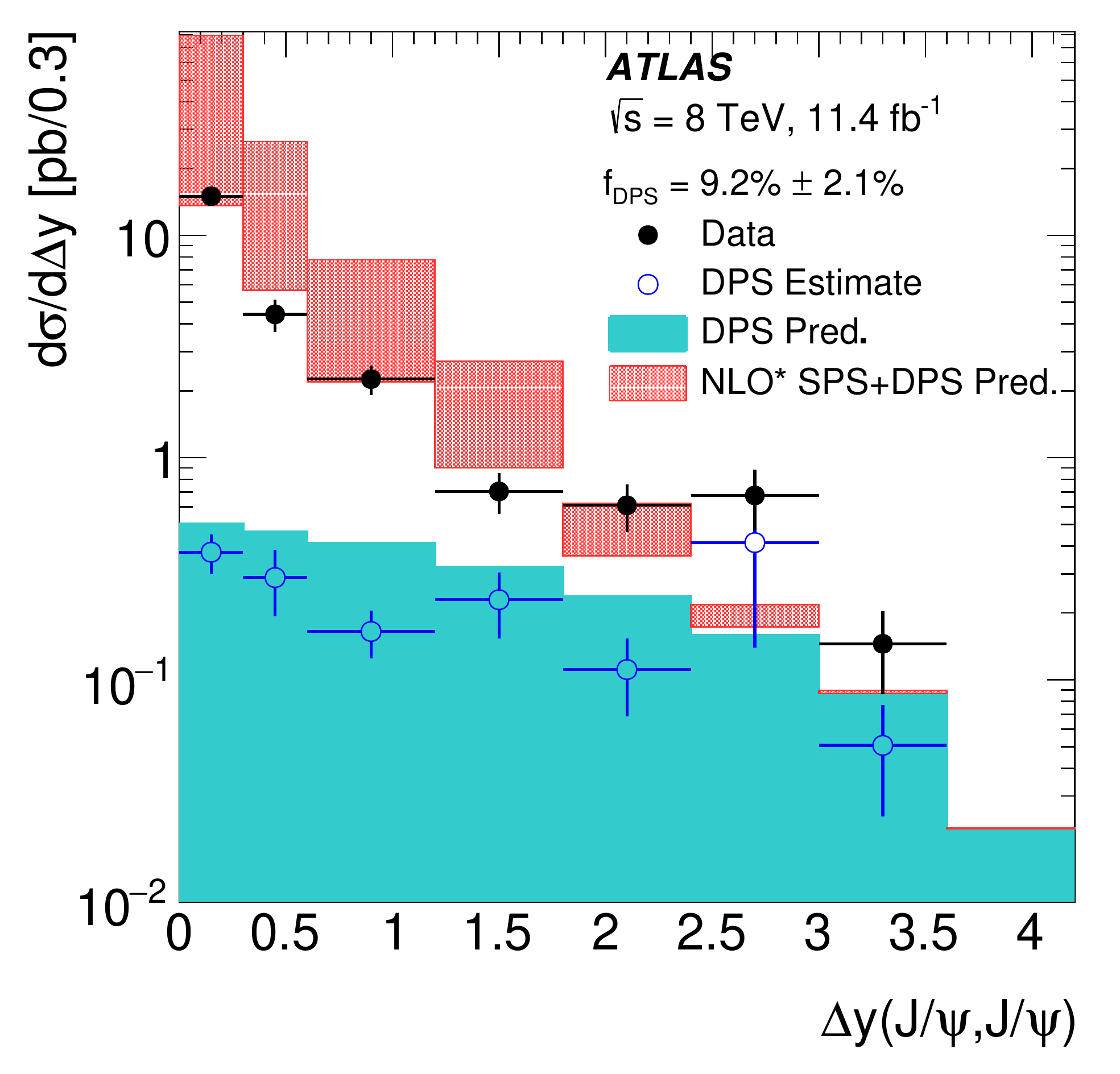}
    }
 \end{picture}
  \caption {\small
    The~DPS and total differential cross sections as 
   a~function of (left)\,the~azimuthal angle between the two \jpsi~mesons, 
   and (right)\,the~difference in rapidity between the two \jpsi~mesons.
   Shown are the~data as well as the~DPS and NLO$^{\ast}$\,SPS predictions.
   The~DPS predictions~\cite{Borschensky:2016nkv}
   are normalized to the~value of $f_{\mathrm{DPS}}$ found in the~data
   and the~NLO$^{\ast}$\,SPS predictions are multiplied by a~constant 
   feed\nobreakdash-down correction factor.  
   The~data\nobreakdash-driven DPS\nobreakdash-weighted distribution and 
   the~total data distribution are compared 
   to the~DPS theory prediction and the~total SPS+DPS prediction.
 }
 \label{fig:dpshf:fig_nine}
\end{figure}

A~measurement with higher centre\nobreakdash-of\nobreakdash-mass 
energy of 8~\tev has been performed by the~ATLAS collaboration.
In~total they observed \mbox{$1160\pm70$} promptly produced \jpsi~pairs.
The~total cross section in the~fiducial region was measured to
be 
\begin{equation*}
\upsigma^{\jpsi\jpsi}=160\pm12\pm14\pm2\pm3\pb,
\end{equation*}
where the first uncertainty is statistical, the second one is
systematic mainly due to the~trigger efficiency determination.
The~third and fourth uncertainties are coming correspondingly
from the known $\jpsi\to\upmu\upmu$ branching fraction
uncertainty and the~luminosity determination uncertainty.
A~data\nobreakdash-driven model\nobreakdash-independent 
technique was used to separate the~DPS and the~SPS contributions.
The overall differential cross section and 
the~differential cross section for DPS~contribution are shown in 
Fig.~\ref{fig:dpshf:fig_nine}.

The~measured DPS\nobreakdash-sensitive distributions were compared
and found in a~good agreement with the~DPS predictions,~\cite{Borschensky:2016nkv}
while  the~SPS distributions have shown some discrepancy
with the~NLO$^*$ predictions~\cite{Lansberg:2014swa,Lansberg:2013qka}
which however could be explained by feed\nobreakdash-down from
higher charmonium states. 
The~DPS fraction was found to be
$\mathrm{f}_{\mathrm{DPS}} = (9.2\pm2.1\stat\pm0.5\syst)\%$, 
resulting in 
\begin{equation*}
  \seff = 6.3\pm1.6\pm1.0\pm0.1\pm0.1\mbarn,
\end{equation*}
where the~first uncertainty is statistical, 
the~second one is
systematic, the~third one comes from the $\jpsi\to\upmu\upmu$
branching fraction and the fourth is an~uncertainty
on the~luminosity determination.

\begin{figure}[t]
  \setlength{\unitlength}{1mm}
  \centering
  \begin{picture}(130,70)
    \put( 0, 0){
      \includegraphics*[width=65mm,height=70mm,%
      ]{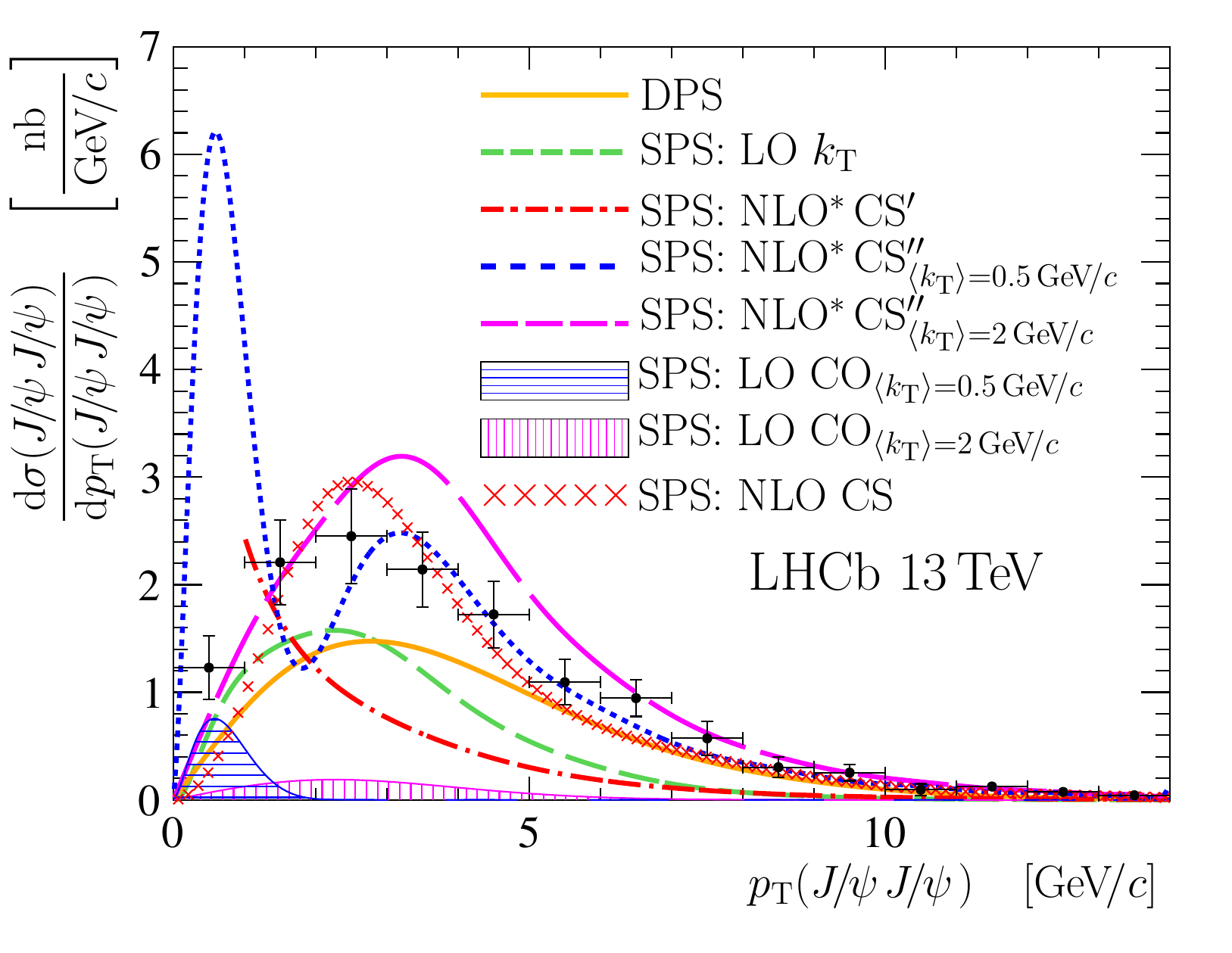}
    }
    \put(65, 0){
      \includegraphics*[width=65mm,height=70mm,%
      ]{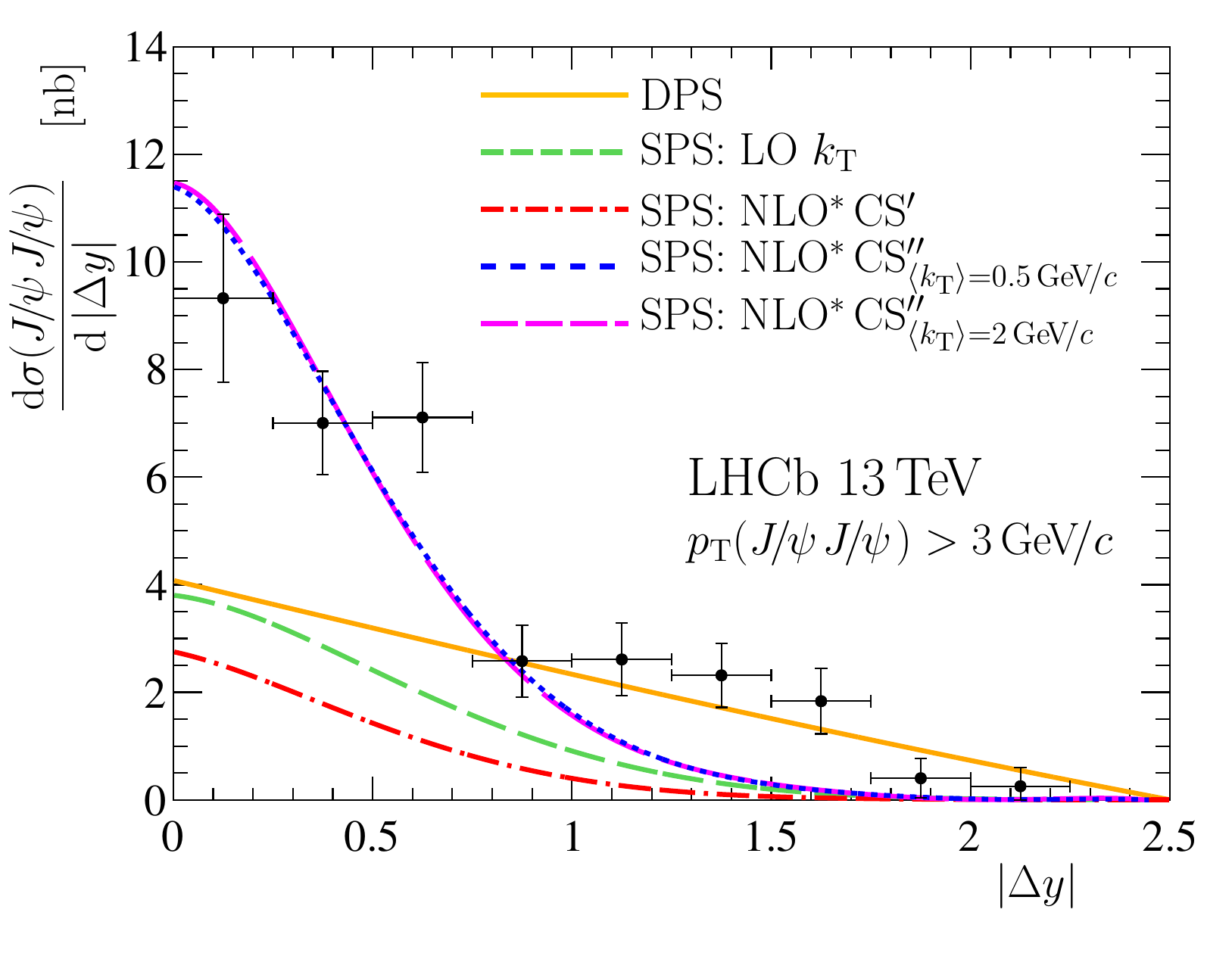}
    }
 \end{picture}
  \caption{\small
    Comparisons between measurements~\cite{Aaij:2016bqq}
   and theoretical predictions~\cite{Lansberg:2014swa,Likhoded:2016zmk,Shao:2012iz,Shao:2015vga,Lansberg:2013qka,Lansberg:2015lva,Sun:2014gca}
   for the~differential cross sections as 
   a~function of (left)~$p_{\mathrm{T}}^{\jpsi\jpsi}$ and 
   (right)~$\left|\Delta y \right|$.
    The~(black) points with error bars represent the~measurements.
    }
 \label{fig:dpshf:fig_ten}
\end{figure}

One~more study of the~double \jpsi~production was performed by
the~LHCb collaboration~\cite{Aaij:2016bqq} at 
the~centre\nobreakdash-of\nobreakdash-mass energy  of~\mbox{$13\tev$}. 
They~observed \mbox{$(1.05\pm0.05)\times10^3$}~signal \jpsi~pairs.
The~total cross section in the~fiducial region
was measured to be 
\begin{equation*}
\upsigma^{\jpsi\jpsi}=15.2\pm1.0\stat\pm0.9\syst\,\mathrm{nb}.
\end{equation*}
A~large statistics of \jpsi~pairs allows to study differential 
cross sections and compare them with different 
theory models. Figure~\ref{fig:dpshf:fig_ten} shows 
the~normalized differential cross sections
as a~function of the~transverse momentum of \jpsi\jpsi~system 
and the~rapidity difference between two \jpsi~mesons.

\begin{figure}[t]
  \setlength{\unitlength}{1mm}
  \centering
  \begin{picture}(130,70)
    \put( 0, 0){
      \includegraphics*[width=65mm,height=70mm,%
      ]{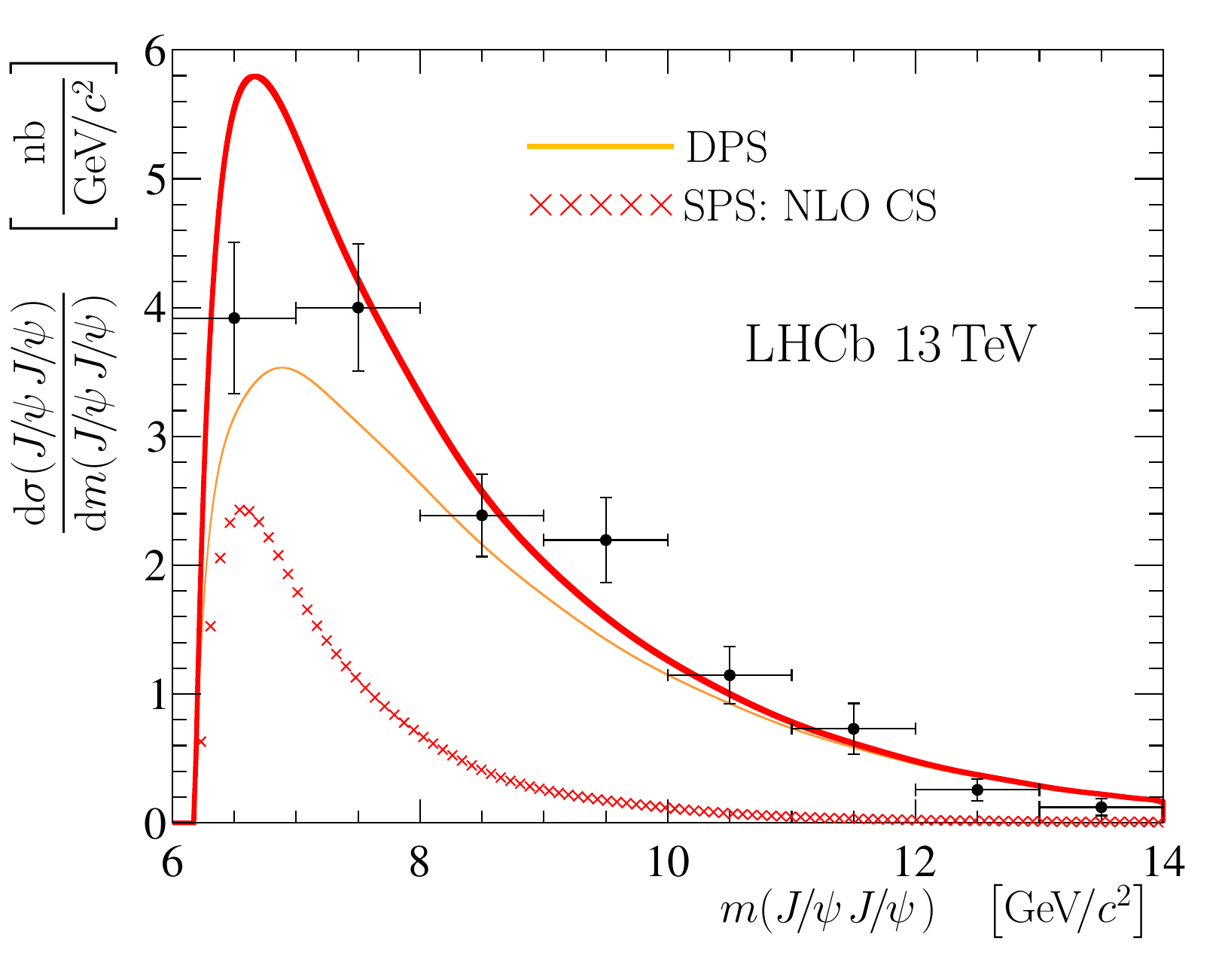}
    }
    \put(65, 0){
      \includegraphics*[width=65mm,height=70mm,%
      ]{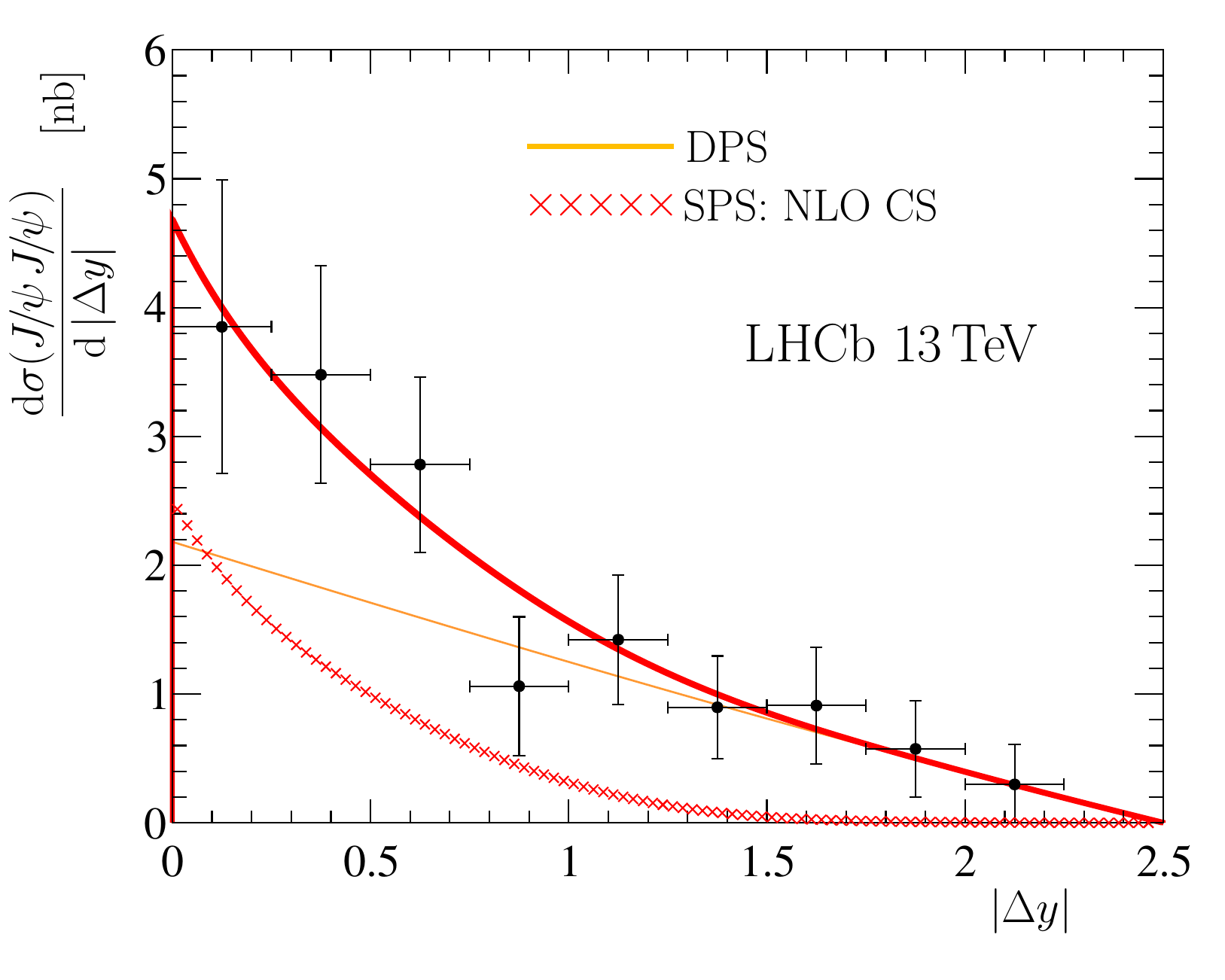}
    }
 \end{picture}
  \caption{\small
    Result of templated DPS fit for 
    $\tfrac{ \deriv\upsigma}{\deriv m }$\,(left) and 
    $\tfrac{ \deriv\upsigma}{\deriv \left| \Delta y \right|}$\,(right). 
    The~points with error bars represent the~data. 
    The~total fit result is shown with the thick\,(red) solid line, 
    the~DPS component is shown with the~thin\,(orange) solid line
    and the~SPS component (full NLO\,CS) is shown with small\,(red) crosses. }
 \label{fig:dpshf:fig_fit_ten}
\end{figure}

The~obtained cross section value is 
interpreted as a~sum of the~SPS and DPS contributions, 
which were separated by a~study of differential
cross sections.
Fit~to the~data was performed with a~function, 
corresponding to the sum of an~SPS and DPS model predictions. 
The~fraction $f_{\mathrm{DPS}}$ was treated as free parameters 
of the~fit.  
Examples of DPS fit to 
$m(\jpsi\jpsi)$ and 
$\left| \Delta y \right|$~distributions
using full NLO\,CS calculations for SPS model
are shown  in~Fig.~\ref{fig:dpshf:fig_fit_ten}. 
For~all fits~the fractions of the~DPS contribution
were found to be higher than~50\%, 
and for many cases compatible with~100\%.  

The~obtained values of $\upsigma_{\mathrm{DPS}}$ 
and the~known~\cite{Aaij:2015rla}  
\jpsi~production cross section
were used to calculate
$\upsigma_{\mathrm{eff}}$.
The~obtained \seff~values are shown in 
Table~\ref{tab:dpshf:tab1} depending on 
the~choice of the~SPS model used.
They~turned out to be slightly higher than 
values  measured from the~central \jpsi~pair production.
In~spite of the~large difference between different SPS models, 
the~values of \seff exhibit only modest model dependency.
 
\begin{table}[htb]
  \centering
  \caption{\small
    Summary of the~\seff~values (in $\!\mbarn$) from DPS~fits~\cite{Aaij:2016bqq} 
    for different SPS models.
    The~uncertainty is statistical~only.
    The~common systematic uncertainty of 12\%, accounting
    for the~systematic uncertainty of $\upsigma\left(\jpsi\jpsi\right)$ and
    the~total uncertainty for $\upsigma(\jpsi)$, is not shown.}
  \begin{tabular*}{1.0\textwidth}{@{\hspace{1mm}}l@{\extracolsep{\fill}}cccc@{\hspace{1mm}}}
    \multirow{2}{*}{Variable}
    & \multirow{2}{*}{LO\,$k_{\mathrm{T}}$~~\cite{PhysRevD.84.054012}}
    & \multicolumn{2}{c}{NLO$^{\ast}$\,CS$^{\prime\prime}$~~\cite{Lansberg:2014swa,Lansberg:2013qka,Lansberg:2015lva,Shao:2012iz,Shao:2015vga}}
    & \multirow{2}{*}{NLO\,CS~~\cite{Sun:2014gca}}
    \\
    &
    & ${\left\langle k_{\mathrm{T}}\right\rangle=2\gevc}$
    & ${\left\langle k_{\mathrm{T}}\right\rangle=0.5\gevc}$
    \\
    \hline
    \\[-1em]
$p^{\jpsi\jpsi}_{\mathrm{T}}$
    & $11.3 \pm 0.6$
    & $10.1 \pm 6.5$
    & $10.9 \pm 1.2$
    & ---
    \\
$y^{\jpsi\jpsi}$
    & ---
    & $11.9 \pm 7.5$
    & $10.0 \pm 5.0$
    & ---
    \\
$m^{\jpsi\jpsi}$
    & $10.6 \pm 1.1$
    & \multicolumn{2}{c}{$10.2 \pm 1.0$}
    & $10.4 \pm 1.0$
    \\
$\left| \Delta y \right|$
    & $12.5 \pm 4.1$
    & $12.2 \pm 3.7$
    & $12.4 \pm 3.9$
    & $11.2 \pm 2.9$
  \end{tabular*}
\label{tab:dpshf:tab1}
\end{table}

\begin{figure}[t]
 \setlength{\unitlength}{1mm}
 \centering
 \begin{picture}(130,70)
   \put( 0, 0){
     \includegraphics*[width=65mm,height=70mm,%
     ]{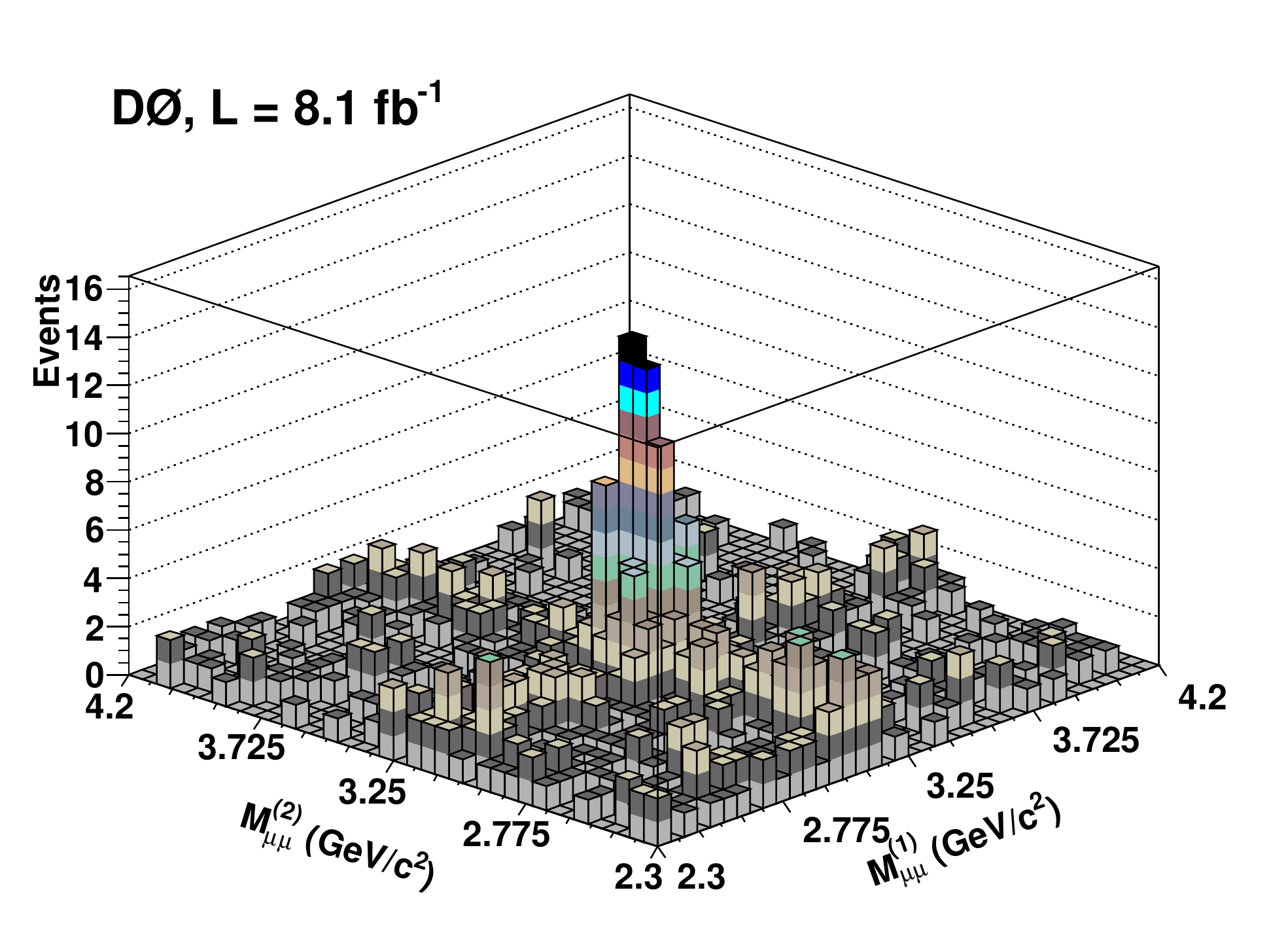}
   }
   \put(65, 0){
     \includegraphics*[width=65mm,height=70mm,%
     ]{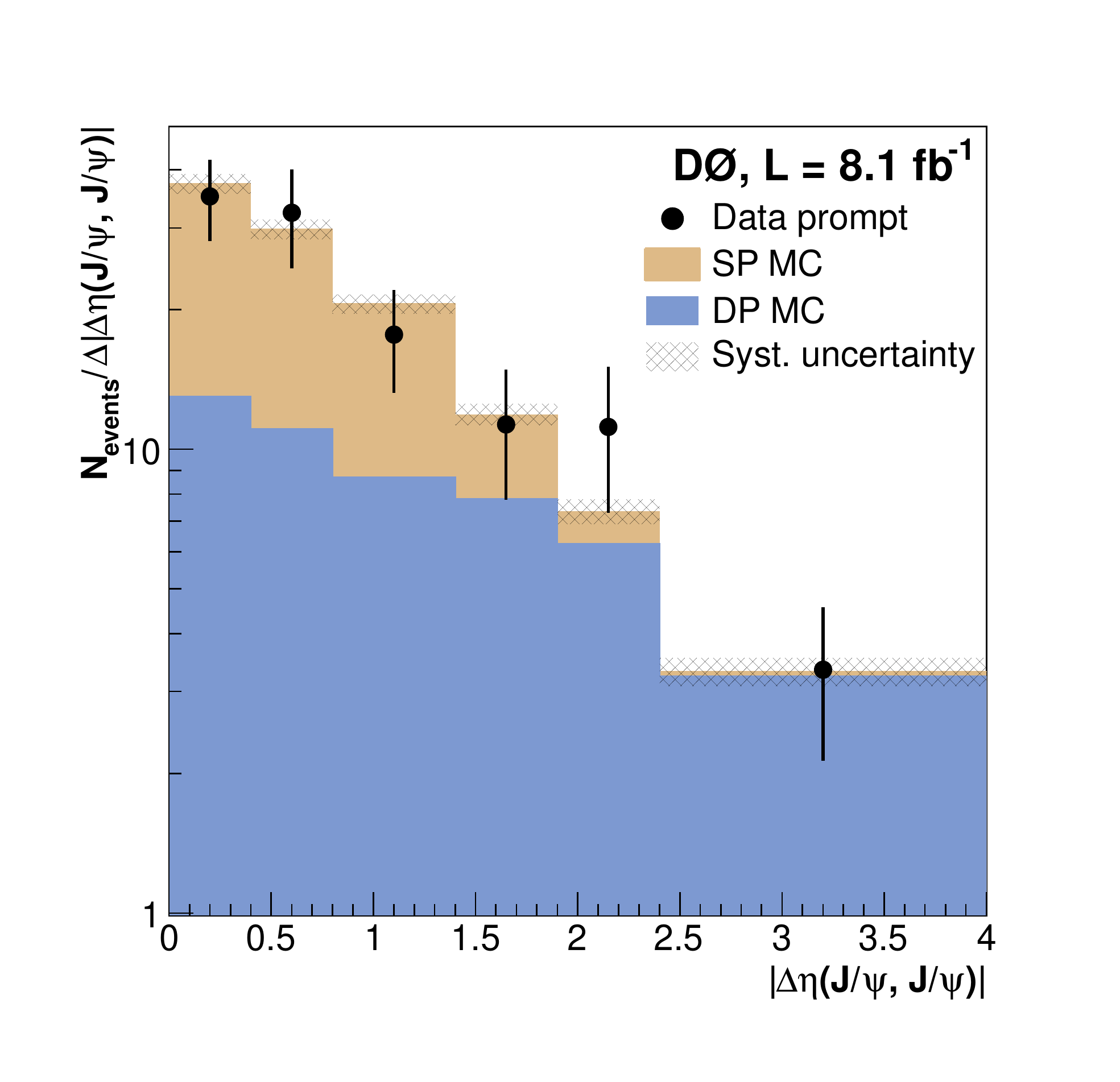}
   }
 \end{picture}
 \caption {\small
   (left)~Dimuon invariant mass distribution in data for two muon pairs.
   (right)~The~distribution of the~rapidity difference between
   two \jpsi candidates in data after background subtraction.
   The distributions for the~SPS and DPS templates are
   shown normalized to their respective fitted fractions. 
   The~uncertainty band corresponds to 
   the~total systematic uncertainty
   on the~sum of SP and DP events.
 }
 \label{fig:dpshf:fig_eleven}
\end{figure}

The D0 collaboration has for the first time observed the~pair
\jpsi~production in $\proton\antiproton$~collisions
at~$\sqs=1.96\tev$ at Tevatron, 
see Fig.~\ref{fig:dpshf:fig_eleven}(left).
\mbox{After}~the~selection 242~pairs of \jpsi~mesons were found, 
about 40\% of 
which were determined to be prompt signal 
\jpsi~pairs. 
The~cross section in 
the~fiducial region
was measured to be 
\begin{equation*}
\upsigma^{\jpsi\jpsi} =  129\pm11\stat\pm37\syst\fb.
\end{equation*}
In~order to study the~SPS and DPS contributions separately,
they were distinguished by fitting the~\mbox{$|\Delta y|$}
distribution by a~sum of SPS and DPS templates with 
the~corresponding fractions used as the~free parameters,
see Fig.~\ref{fig:dpshf:fig_eleven}(right). 
The~DPS fraction $\mathrm{f}_{\mathrm{DPS}}$ 
was found to be~\mbox{$(42\pm12)\%$}, 
leading to the following values of the~cross sections:
\mbox{$59\pm6\stat\pm22\syst\fb$} and 
\mbox{$70\pm6\stat\pm22\syst\fb$} for SPS and DPS processes, 
respectively.
The~corresponding \seff  was measured to be
\mbox{$4.8\pm0.5\stat\pm2.5\syst\mbarn$}.

The~measurements of  associated productions 
involving bottomonia resonances were performed by 
D0~and CMS~collaborations.
The~D0 collaboration observed 
the~associated
\jpsi and $\PUpsilon$~production~\cite{Abazov:2015fbl}
using 8~$\mathrm{fb}^{-1}$ of data collected at~\mbox{$\sqs=1.96~\tev$} 
in $\proton\antiproton$~collisions.
This~combination is expected to be produced 
predominantly through DPS mechanism with SPS contribution
suppressed~\cite{Baranov:2011ch,Lansberg:2015lva}
by an~additional powers of $\upalpha_{\mathrm{s}}$.
It results in the~hierarchy of expected SPS
production cross sections
\mbox{$\upsigma^{\jpsi\PUpsilon}_{\mathrm{SPS}}    < 
\upsigma^{\PUpsilon\PUpsilon}_{\mathrm{SPS}} < 
\upsigma^{\jpsi\jpsi}_{\mathrm{SPS}}$}
that is opposite to more intuitive DPS hierarchy 
\mbox{$\upsigma^{\PUpsilon\PUpsilon}_{\mathrm{DPS}} < 
       \upsigma^{\jpsi\PUpsilon}_{\mathrm{DPS}}     < 
       \upsigma^{\jpsi\jpsi}_{\mathrm{DPS}}$}.
That~makes the~observation of \jpsi and $\PUpsilon$~pair
production and the~precise measurement of the~hierarchy
of the~production cross sections very important.

\begin{figure}[t]
  \setlength{\unitlength}{1mm}
  \centering
  \begin{picture}(130,70)
    \put( 0, 0){
      \includegraphics*[width=65mm,height=70mm,%
      ]{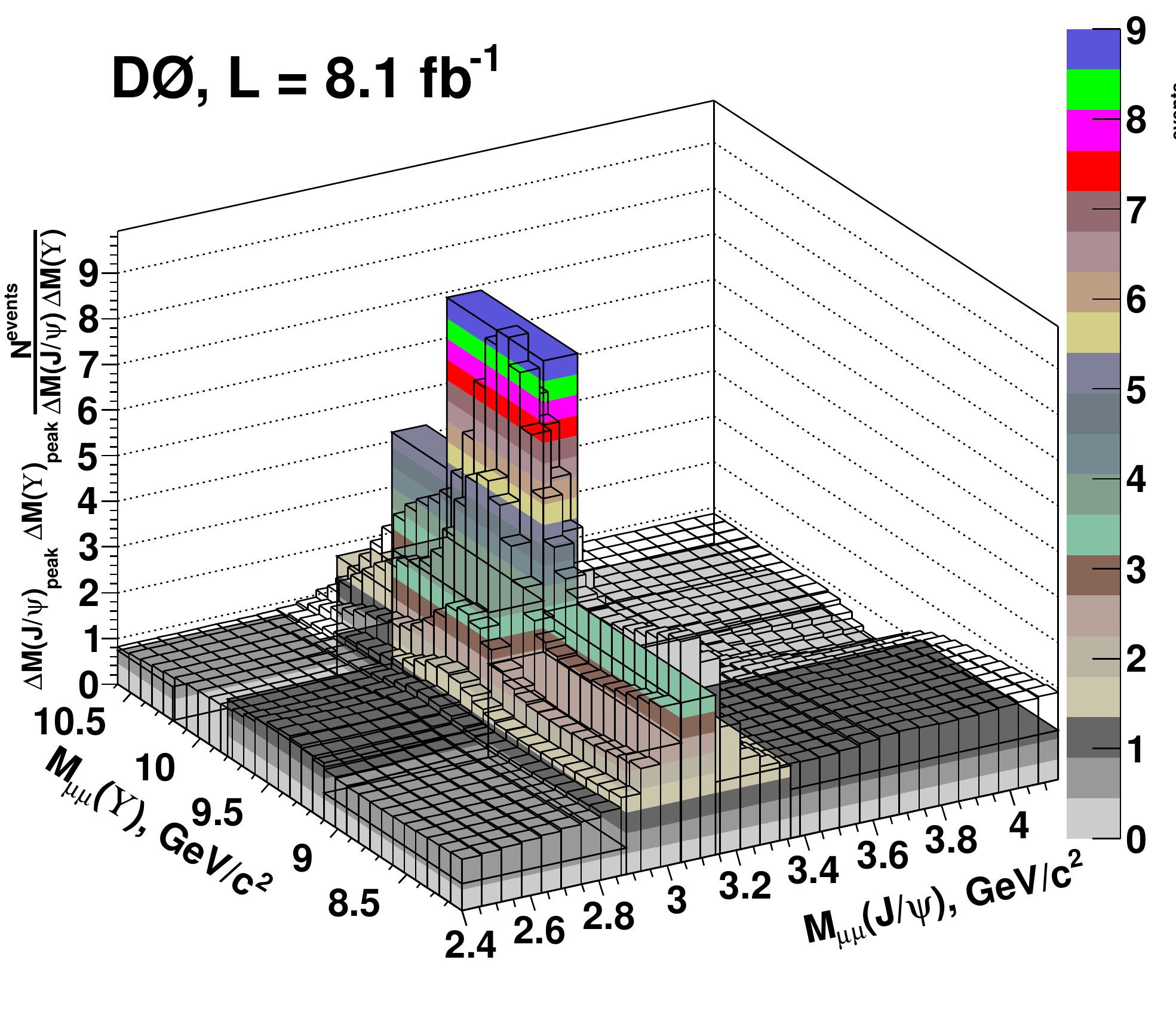}
    }
    \put(65, 0){
      \includegraphics*[width=65mm,height=70mm,%
      ]{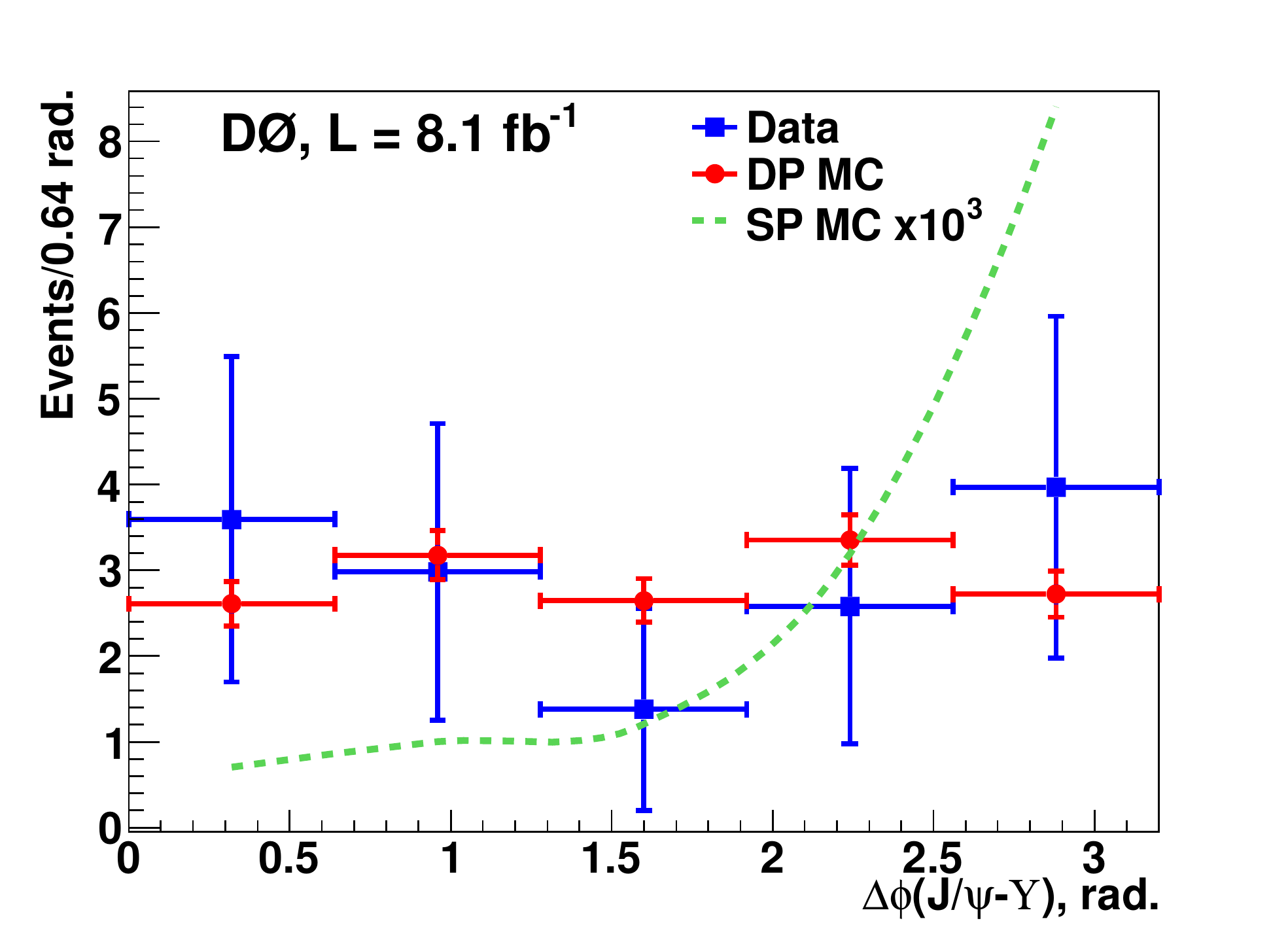}
    }
 \end{picture}
  \caption {\small
    (left)~Dimuon invariant mass distribution in data  for two muon pairs
   together with the~two\nobreakdash-dimensional fit surface. 
   The~scaling factor is applied so that the~height of the~peak bin 
   is the~number of observed events in that~bin.
   (right)~The~distribution of the~azimuthal angle between 
   the~\jpsi and $\PUpsilon$ candidates in data
   after background subtraction.
   Also shown the~expectations from DPS and SPS processes in arbitrary units.
 }
 \label{fig:dpshf:fig12}
\end{figure}

Dimuon invariant mass distribution in data  for 
two muon pairs in~the~fiducial region 
of $p_{\mathrm{T}}^{\upmu}>2\gevc$ and
$\left|\Peta^{\upmu}\right|<2$ is shown 
in~Fig.~\ref{fig:dpshf:fig12}(left) together with 
the~two\nobreakdash-dimensional fit surface. 
In~total \mbox{$12\pm3.8\stat\pm2.8\syst$} 
promptly
produced $\jpsi\PUpsilon$~pairs are observed,
corresponding 
to the~significance of~$3.2\sigma$. 

The~distribution of azimuthal
angle between \jpsi and $\PUpsilon$ mesons 
is presented in Fig.\ref{fig:dpshf:fig12}(right).
This~distribution showed a~good agreement with
the~dominance of DPS mechanism.
It~allows to translate the~measured production 
cross section for $\jpsi\PUpsilon$~pairs 
\begin{equation*}
\upsigma^{\jpsi\PUpsilon}= 27\pm9\stat\pm7\syst\fb
\end{equation*}
into the~effective cross section
\mbox{$\seff=2.2\pm0.7\stat\pm0.9\syst\mbarn$}.
The~obtained value of~\seff 
is below the~previous measurements involving
heavy quarkonium. 

\begin{figure}[t]
  \setlength{\unitlength}{1mm}
  \centering
  \begin{picture}(130,90)
    \put( 0,15){
      \includegraphics*[width=65mm,height=55mm,%
      ]{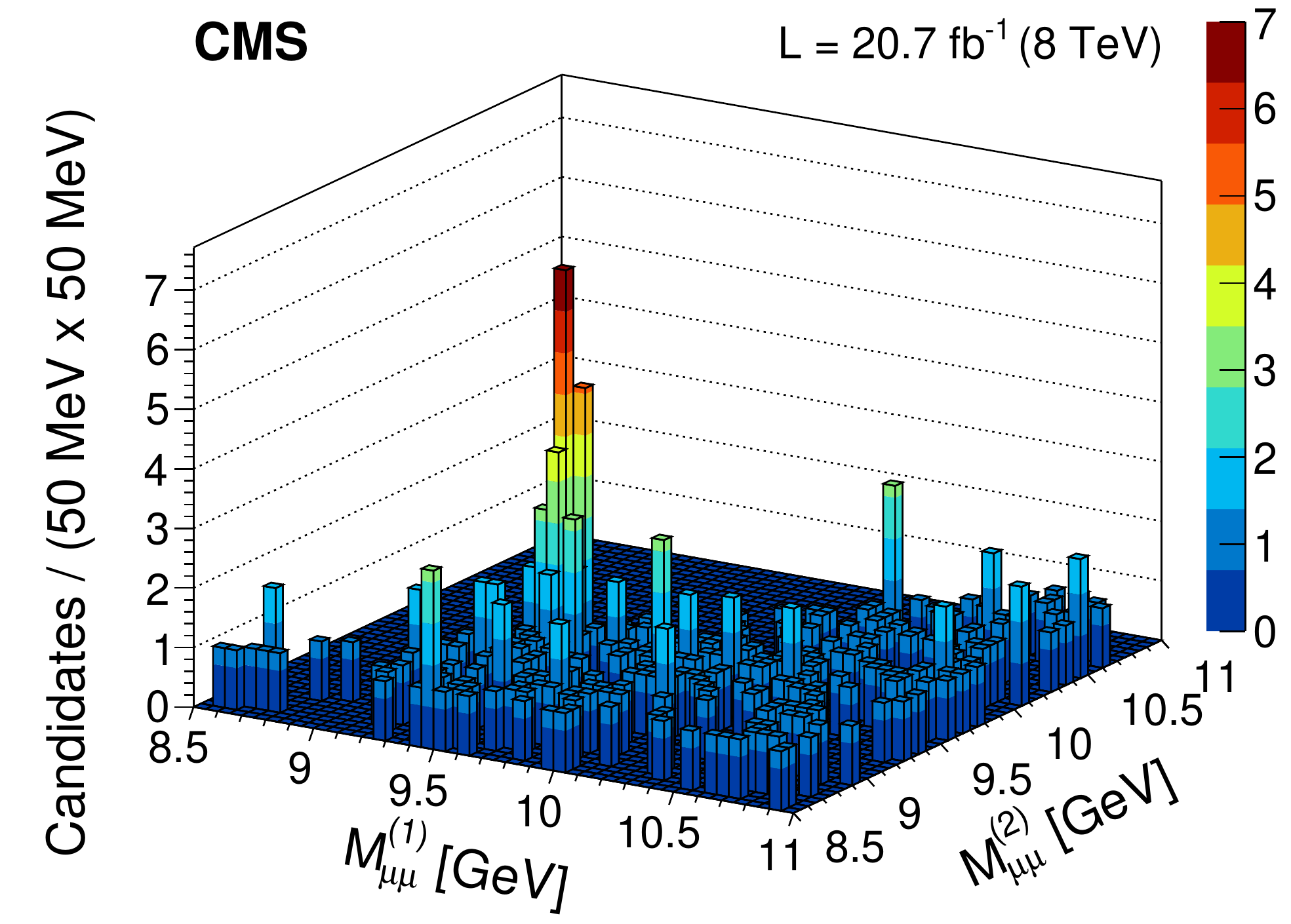}
    }
    \put(65, 0){
      \includegraphics*[width=65mm,height=45mm,%
      ]{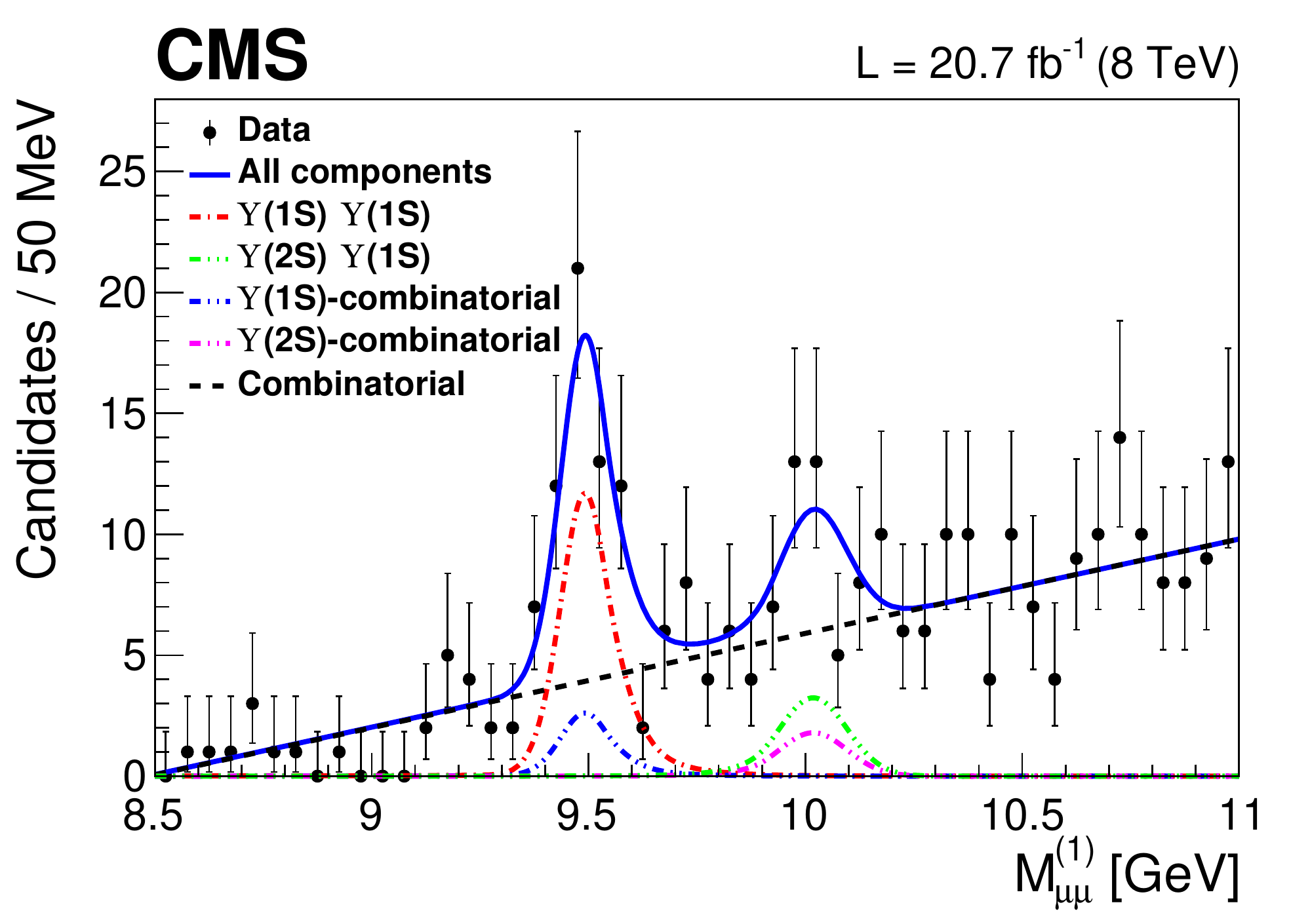}
    }
    \put(65,45){
      \includegraphics*[width=65mm,height=45mm,%
      ]{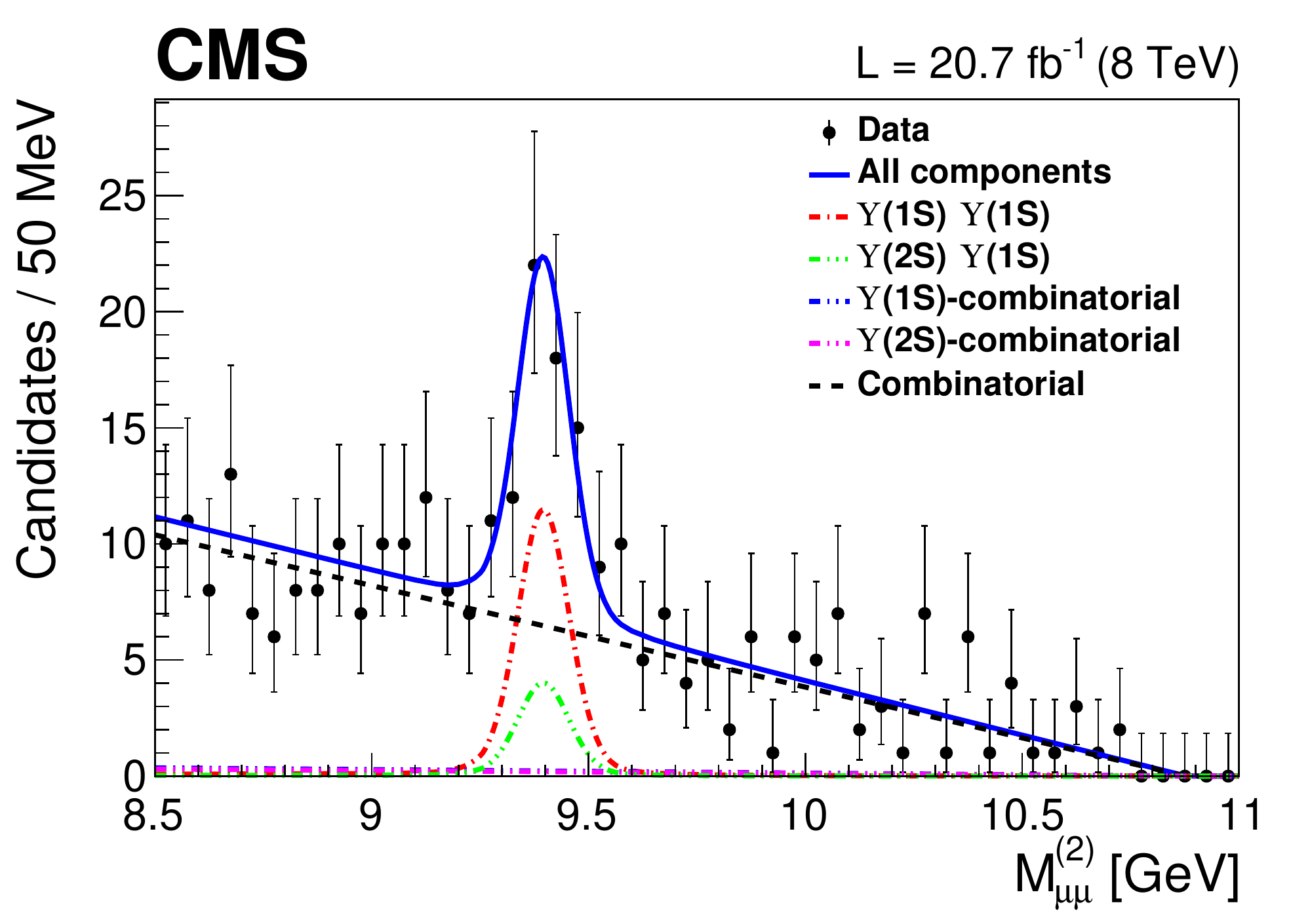}
    }
 \end{picture}
  \caption {\small
    (left)~Dimuon invariant mass distribution in data  for two muon pairs.
   (right)~Invariant mass distributions of 
   the~lower-mass muon pair\,(top) and  
   the~higher-mass muon pair\,(bottom). 
   The~data are shown by the~points. 
   The~different curves show the~contributions 
   of the~various event categories from the~fit.
 }
 \label{fig:dpshf:fig13}
\end{figure}

The~CMS collaboration made a~first 
observation~\cite{Khachatryan:2016ydm}
of double bottomonium production using $20.7\invfb$ of data
collected in $\proton\proton$~collisions 
at~centre\nobreakdash-of\nobreakdash-mass
energy of~\mbox{8\tev}.
The~dimuon invariant mass distribution in data 
for two muon pairs in~the~fiducial region of 
$\left| y^{\PUpsilon}\right| < 2$ 
is presented in Fig.~\ref{fig:dpshf:fig13}(left).
A~signal yield of $38\pm7$ pairs of $\PUpsilon\mathrm{(1S)}$~mesons 
is determined from two\nobreakdash-dimensional fit.
The~projections of two\nobreakdash-dimensional fit
are shown in Fig.~\ref{fig:dpshf:fig13}(right).
The~significance of the signal exceeds~$5\sigma$.

The~production cross section of 
$\PUpsilon\mathrm{(1S)}$~pairs was 
measured to  be
\begin{equation*}
\upsigma^{\PUpsilon\PUpsilon}=68.8\pm12.7\pm7.4\pm2.8\pb,
\end{equation*}
where the~first uncertainty is statistical,
the~second one is systematic and the~third one 
comes from the~known  value of $\PUpsilon\to\mumu$~
branching fraction.
The~DPS contribution in this channel is expected to be
small~\cite{Novoselov:2011ff} relative to the~SPS one.
Using a~conservative estimate of 
\mbox{$\mathrm{f_{DPS}} = 10\%$},
the~effective cross section was 
estimated to be $6.6\mbarn$. On~the~other hand
the~SPS prediction~\cite{Berezhnoy:2012tu}
with feed\nobreakdash-down 
from higher states gives a~value of
\mbox{$\upsigma^{\PUpsilon\PUpsilon}_{\mathrm{SPS}} = 48\pb$}
which combined with the~$\upsigma^{\PUpsilon\PUpsilon}$ 
measured in data leads to 
\mbox{$\mathrm{f_{DPS}} = 30\%$} and, correspondingly 
\mbox{$\upsigma_{\mathrm{eff}} = 2.2~\mathrm{mb}$}.
Both~estimates are in line with values obtained from
other heavy quarkonium measurements. Higher statistics
would allow to make a~more precise conclusions.

\section{Other DPS measurements involving heavy quarks}

Besides the~studies described above, 
there are other processes
involving heavy quarks which can tell a~lot about charm parton
distribution inside the proton, charm production mechanism
and the DPS. Good~examples of such studies are measurements of
an~associated production of charm and a~\W or a~\Z~boson.
It~should be noted that for~these final states one probes 
the~PDFs in the~region of relatively high-$x$ partons, 
where  significant violation~\cite{Korotkikh:2004bz,Gaunt:2010pi} 
of factorization hypothesis is expected.

The~CMS and ATLAS collaborations studied~\cite{Chatrchyan:2013uja,Aad:2014xca}
an~associated production
of a~\W~boson with open\nobreakdash-charm considering 
both a~case of explicitly
produced \D~meson and 
a~case of jet initiated by the \cquark~quark.
Each~experiment used about~\mbox{$5\invfb$} of data collected
in $\proton\proton$~collisions at the~centre\nobreakdash-of\nobreakdash-mass 
energy of~\mbox{$\sqs= 7\tev$}. 
The~total and differential cross sections
of these processes were measured and found to be consistent between
the~two experiments and with theoretical predictions.


The~LHCb experiment observed~\cite{Aaij:2014hea} 
the~associated production of a~\Z~boson and an~open\nobreakdash-charm
meson\,(\Dz or \Dp)
in the~forward region using data collected at\mbox{$\sqs = 7\tev$}.
Seven candidate events for associated production of a~\Z~boson 
with a~\Dz~meson and four candidate events for a~\Z~boson with 
a~\Dp~meson were observed with a~combined significance of 
$5.1\sigma$. The~fiducial cross sections of these 
processes were measured,  but the~lack of statistics didn't 
allow to disentangle the~DPS and SPS contributions.

With higher statistics these studies promise to give more interesting
and useful information for understanding the heavy quark production
mechanisms and in particular the DPS studies.

\section{Summary}\label{sec:dpshf:summary}

\begin{figure}[htb]
  \centering
  \setlength{\unitlength}{1mm}
  \begin{picture}(130,165) 
    \put(-25,-10){
      \includegraphics*[width=165mm,height=175mm,%
      ]{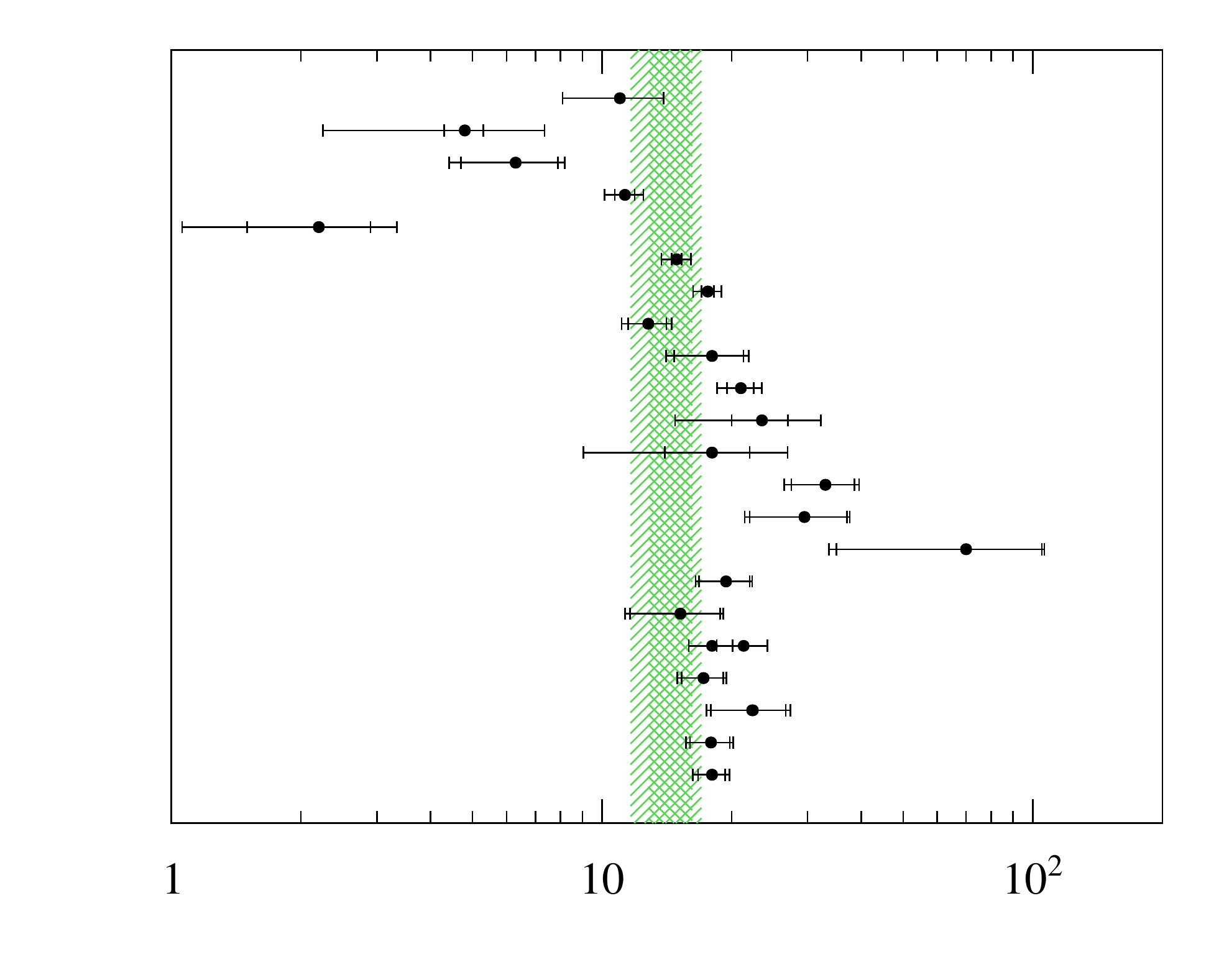}
    }
    \put( 55,0){ \Large $\upsigma_{\mathrm{eff}}$}
    \put(120,0){ \Large $\left[\!\mbarn\right]$}
    \put(75,146.0) {\scriptsize$\jpsi\jpsi$\ CMS~\cite{Khachatryan:2014iia}+LS~\cite{Lansberg:2014swa}\ $\proton\proton @\sqs=7\tev$} 
    \put(75,140.5) {\scriptsize$\jpsi\jpsi$\ D0~\cite{Abazov:2014qba}\ $\proton\antiproton @\sqs=1.96\tev$} 
    \put(75,135.0) {\scriptsize$\jpsi\jpsi$\ ATLAS~\cite{Aaboud:2016fzt}\ $\proton\proton @\sqs=8\tev$} 
    \put(75,129.5) {\scriptsize$\jpsi\jpsi$\ LHCb~\cite{Aaij:2016bqq}\ $\proton\proton @\sqs=13\tev$} 
    \put(75,124.0) {\scriptsize$\PUpsilon\jpsi$\ D0~\cite{Abazov:2015fbl}\ $\proton\antiproton @\sqs=1.96\tev$} 
    \put(3,118.0)  {\scriptsize$\jpsi\Dz$\ LHCb~\cite{Aaij:2012dz}\ $\proton\proton @\sqs=7\tev$} 
    \put(3,112.5)  {\scriptsize$\jpsi\Dp$\ LHCb~\cite{Aaij:2012dz}\ $\proton\proton @\sqs=7\tev$} 
    \put(3,107.0)  {\scriptsize$\jpsi\Ds$\ LHCb~\cite{Aaij:2012dz}\ $\proton\proton @\sqs=7\tev$} 
    \put(3,101.5)  {\scriptsize$\jpsi\Lc$\ LHCb~\cite{Aaij:2012dz}\ $\proton\proton @\sqs=7\tev$} 
    \put(3, 94.0)  {\scriptsize$\DzDz$\ LHCb~\cite{Aaij:2012dz}\ $\proton\proton @\sqs=7\tev$} 
    \put(3, 88.5)  {\scriptsize$\DzDp$\ LHCb~\cite{Aaij:2012dz}\ $\proton\proton @\sqs=7\tev$} 
    \put(3, 83.0)  {\scriptsize$\DzDs$\ LHCb~\cite{Aaij:2012dz}\ $\proton\proton @\sqs=7\tev$} 
    \put(3, 77.5)  {\scriptsize$\DpDp$\ LHCb~\cite{Aaij:2012dz}\ $\proton\proton @\sqs=7\tev$} 
    \put(3, 72.0)  {\scriptsize$\DpDs$\ LHCb~\cite{Aaij:2012dz}\ $\proton\proton @\sqs=7\tev$} 
    \put(3, 66.5)  {\scriptsize$\DpLc$\ LHCb~\cite{Aaij:2012dz}\ $\proton\proton @\sqs=7\tev$} 
    \put(3, 60.0)  {\scriptsize$\PUpsilon\mathrm{(1S)}\Dz$\ LHCb~\cite{Aaij:2015wpa}\ $\proton\proton @\sqs=7\tev$} 
    \put(3, 54.5)  {\scriptsize$\PUpsilon\mathrm{(1S)}\Dp$\ LHCb~\cite{Aaij:2015wpa}\ $\proton\proton @\sqs=7\tev$} 
    \put(3, 49.0)  {\scriptsize$\PUpsilon\mathrm{(1S)}\D^{0,+}$\ LHCb~\cite{Aaij:2015wpa}\ $\proton\proton @\sqs=7\tev$}  
    \put(3, 43.5)  {\scriptsize$\PUpsilon\mathrm{(1S)}\Dz$\ LHCb~\cite{Aaij:2015wpa}\ $\proton\proton @\sqs=8\tev$} 
    \put(3, 38.0)  {\scriptsize$\PUpsilon\mathrm{(1S)}\Dp$\ LHCb~\cite{Aaij:2015wpa}\ $\proton\proton @\sqs=8\tev$} 
    \put(3, 32.5)  {\scriptsize$\PUpsilon\mathrm{(1S)}\D^{0,+}$\ LHCb~\cite{Aaij:2015wpa}\ $\proton\proton @\sqs=8\tev$}  
    \put(3, 27.0)  {\scriptsize$\PUpsilon\mathrm{(1S)}\D^{0,+}$\ LHCb~\cite{Aaij:2015wpa}\ $\proton\proton @\sqs=7\&8\tev$} 
  \end{picture}
  \caption {\small
    Summary of \seff measurements with heavy quarks.
    The~inner error bars indicate 
    the~statistical uncertainty whilst 
    the~outer error bars indicate 
    the~sum of statistical and systematic uncertainties 
    in quadrature.
    The~hatched area shows the~reference value~\cite{Abe:1997xk} 
    of \mbox{$\seff=14.5\pm1.7^{+1.7}_{-2.3}\mbarn$}
    measured in multi\nobreakdash-jet events at the~Tevatron.
  }\label{fig:dpshf:fig_last}
\end{figure}
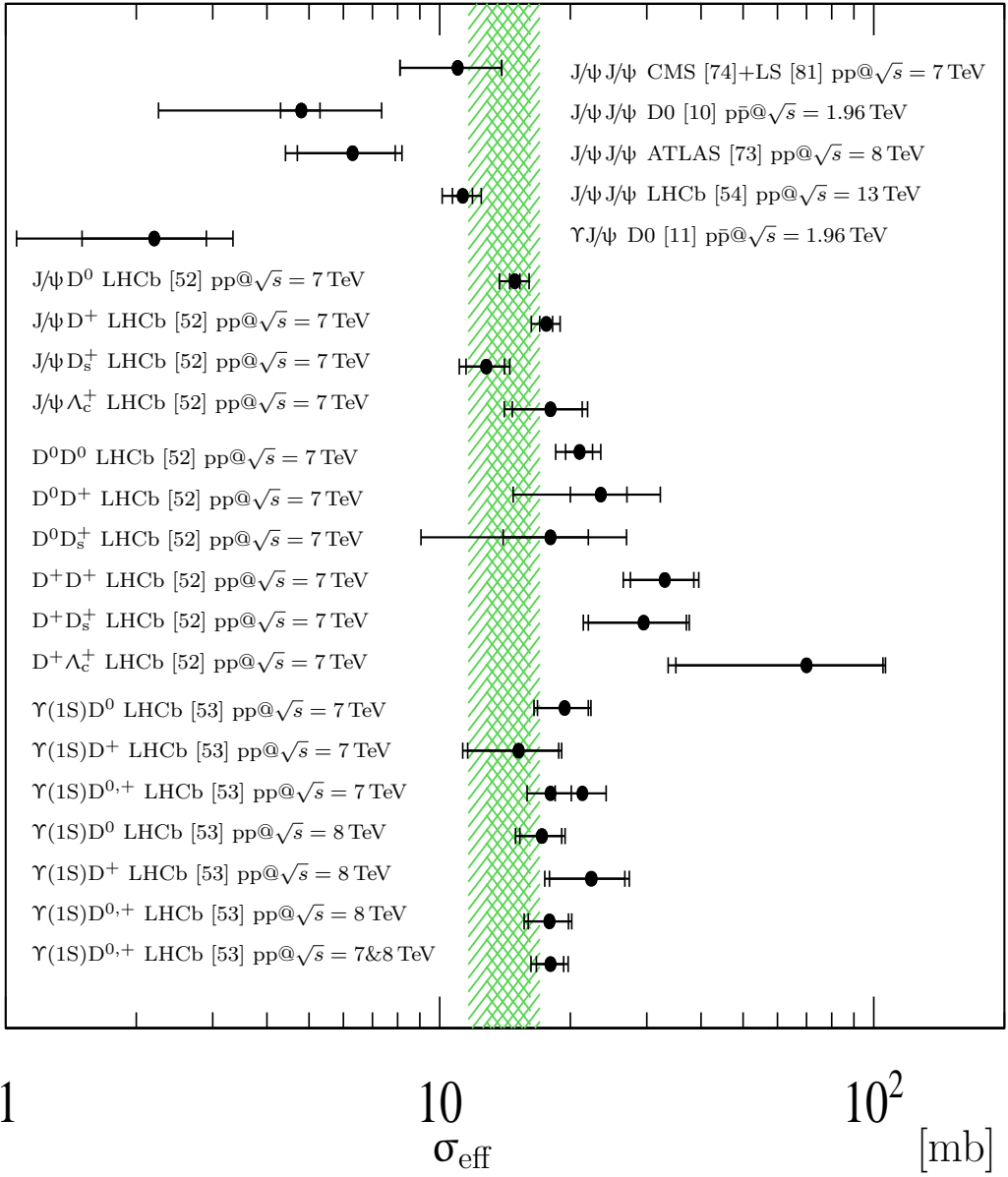

Nowadays a~study of DPS using heavy quarks provides 
the~most precise determination of the~parameter~\seff. 
Figure~\ref{fig:dpshf:fig_last} summarizes all available 
measurements of \seff with heavy quarks.

There are no clear patterns in the~measured values of \seff, 
however the~values of \seff parameter measured with 
the~double quarkonia final state 
are a~bit lower than the~reference value~\cite{Abe:1997xk} 
of \mbox{$\seff=14.5\pm1.7^{+1.7}_{-2.3}\mbarn$}
measured in multi\nobreakdash-jet events at the~Tevatron.
This~could be a~sign that the~spatial region
occupied by gluons within the~proton is smaller 
than that occupied by quarks.
On~the~other hand, 
the~measurements~\cite{Aaij:2012dz} 
with two open\nobreakdash-flavour hadrons give 
values of \seff larger than the~reference value.
These~values are in very good 
agreement~\cite{Blok:2016lmd} 
with calculations.
The~measurements with quarkonia and open\nobreakdash-flavour 
hadrons are well consistent with the~reference value of \seff
and all other measurements~\cite{Bansal:2014paa,Aaboud:2016dea}
of \seff using multi\nobreakdash-jet, 
di\nobreakdash-jet+\W and 
$\gamma$+3\nobreakdash-jets processes.
It~should be noted that for many measurements 
the~uncertainties are large,  
and formally only one~measurement~\cite{Abazov:2015fbl} 
is not consistent 
with the~reference value of~\seff. 
Better precision is needed to~conclude on universality of \seff.
For~large part of measurements, especially for production 
cross section of $\PUpsilon\PUpsilon$ and $\jpsi\PUpsilon$~pairs, 
the~precision 
is limited by the~statistics.
Currently none of the~LHC experiments used 
the~full Run\nobreakdash-I data sample for DPS studies.
Extending these analyzes to full Run\nobreakdash-I and subsequent 
larger Run\nobreakdash-II data sets will allow significant improvement of 
the~precision of \seff~measurements and it will allow 
a~definite conclusion 
on the~universality of \seff. 
Also~the~analysis of larger data sets will allow 
to study the~DPS~processes with 
open\nobreakdash-beauty hadrons in the~final state.
A~huge statistics of events with pairs 
of open\nobreakdash-charm hadrons at LHCb 
could allow the~precise measurement of \seff~separately 
in the~different kinematic regions, 
providing the ultimate test for 
the~universality of \seff. 
It~is worth to mention that for~the~precise measurements 
of DPS processes in $\proton\proton$~collisions at $\sqs=13\tev$ 
one will need to account for the~contribution from 
the~triple parton scattering process~\cite{Maciula:2017meb}.

\clearpage

\bibliographystyle{LHCb}
\bibliography{dps-hf}


\end{document}